\documentclass[aps, nofootinbib,superscriptaddress, preprintnumbers, longbibliography]{article}
\usepackage{tikz}
\usepackage{jcappub}
\usepackage[utf8]{inputenc}
\usepackage{amsmath,amsfonts,amssymb}
\usepackage{graphicx}
\usepackage[toc,page]{appendix}
\usepackage{cleveref}
\usepackage[loose]{units}
\usepackage[markup=default]{changes}
\definechangesauthor[name={Red}, color=red]{YN}
\setdeletedmarkup{\textcolor{red}{\sout{#1}}}
\usepackage{booktabs}
\usepackage{mathtools}
\def\la{~\mbox{\raisebox{-.6ex}{$\stackrel{<}{\sim}$}}~}
\def\ga{~\mbox{\raisebox{-.6ex}{$\stackrel{>}{\sim}$}}~}

\begin{document}

\title{
Analytic results in aligned axion inflation
}

\author[a,b]{Federico Greco,}
\author[a,b]{Marco Peloso}

\affiliation[a]{Dipartimento di Fisica e Astronomia “G. Galilei”, Universit\`a degli Studi di Padova, via Marzolo 8, I-35131 Padova, Italy}
\affiliation[b]{INFN, Sezione di Padova, via Marzolo 8, I-35131 Padova, Italy}

\emailAdd{federico.greco@pd.infn.it, marco.peloso@pd.infn.it}

\abstract{The original model of axion natural inflation produces a tensor-to-scalar ratio above the experimental limit. Aligned axion inflation admits inflationary trajectories that originate near a saddle point of the two-field potential, and terminate due to the instability of the orthogonal direction. The phenomenology of these solutions is within the current constraints, and a range of parameters will be probed by the next stage CMB experiments. We provide the analytic solution for these trajectories and very compact analytic expressions for the associated phenomenology. For parameters leading to the observed value for the scalar spectral tilt the extension of the inflationary trajectory is sub-Planckian. However, one eigenvalue of the axion kinetic matrix (in the basis that diagonalizes the potential) is trans-Planckian. Finally, we discuss the post-inflationary evolution after the instability. In some cases, the fields reach a second inflationary valley, connected to a minimum. Multiple stages of inflation might be a more general occurrence in multiple-field inflationary models with trajectories starting next to critical points.
}

\maketitle

\section{Introduction}
\label{sec:intro}

Cosmological inflation~\cite{Guth:1980zm,Linde:1982uu,Albrecht:1982wi} provides initial conditions for the early universe cosmology in excellent agreement with large-scale measurements~\cite{Planck:2018jri,BICEP:2021xfz}. However, the particle physics realization of inflation remains unknown, and faces significant theoretical challenges. This is particularly true for large-field models of inflation, namely for those models in which the inflaton scans a range $\Delta \phi$ of order of the Planck scale $M_p \equiv 1/ \sqrt{8 \pi G_N}$ (where $G_N$ is the Newton constant) during inflation, that own their popularity also to the fact that they might produce an observable tensor signal~\cite{Lyth:1996im}. For such models, any irrelevant operator $\Delta V = \frac{c_n \, \phi^{n+4}}{M^n}$, where $M$ is a mass scale and $c_n$ a  dimensionless coefficient, generally spoils the required flatness of the potential over the range $\Delta \phi$, unless $M > M_p$ or all the coefficients are tuned to  values much smaller than one.

This problem might be solved~\cite{Freese:1990rb,Adams:1992bn} if the model enjoys a shift symmetry that forbids terms that are not invariant under $\phi \to \phi + c$, where $c$ is a constant. A typical example is that of axion-like particles (ALPs), for which the shift symmetry can be broken by instantons or other non-perturbative string effects to a discrete subgroup $\phi \rightarrow \phi + 2 \pi f$, where $f$ is a mass dimension $1$ quantity denoted as the axion decay constant, allowing for a potential term for the inflaton-axion. This breaking is technically natural, since any radiative correction must be proportional to the tree-level potential in order to also break the shift symmetry and generate a potential term. The simplest potential that satisfies this  residual symmetry was employed in the original model~\cite{Freese:1990rb}
\begin{equation}
    V(\phi)= \Lambda^4 \left[ 1- \cos \left(\frac{\phi}{f} \right) \right],
\end{equation}
which is known as \textit{natural inflation}. Despite this model is now ruled out by CMB observations~\cite{Planck:2018jri}, the idea of~\cite{Freese:1990rb} has originated a widely studied class of inflationary models known as axion inflation~\cite{Pajer:2013fsa}. To be compatible with CMB measurements, and to generate a potentially observable tensor signal, such models typically require a trans-Planckian decay constant. This constitutes a problem since quantum gravity effects are expected to break the shift symmetry of the axion as any global symmetry \cite{Ghigna:1992iv,Holman:1992us,Kamionkowski:1992mf,Giddings:1987cg,Giddings:1989bq}. This issue does not arise if we deal with string theory axions, since in that case the shift symmetry originates from an underlying gauge symmetry which is protected against quantum gravity corrections \cite{Kallosh:1995hi}. However, no fully controlled constructions exhibiting trans-Planckian decay constants in string theory axions are known \cite{Banks:2003sx}.

Several mechanisms proposed to overcome these difficulties have appeared in the literature, including \textit{N-flation}~\cite{Dimopoulos:2005ac}, where inflation is due to the simultaneous presence of many sub-Planckian axions (some challenges towards string realizations of this idea are discussed in \cite{Rudelius:2014wla}), and \textit{axion monodromy}~\cite{McAllister:2008hb}, characterized by a monodromy that extends the field
range of a string axion to beyond the Planck-scale. An alternative idea is to modify the evolution of an axion-inflaton exploiting a coupling to gauge fields \cite{Anber:2009ua,Adshead:2012kp}. An earlier proposal is that of \textit{aligned axion inflation}~\cite{Kim:2004rp}, which is characterized by two axion fields and potential terms that involve nearly the same linear combination of the two axions. In this way, the orthogonal combination becomes extremely flat, giving rise to an effective axion decay constant, and an inflaton excursion, much greater than the one suggested by the individual scales present in the two terms. 

The alignment mechanism of~\cite{Kim:2004rp} has been the object of a number of studies, and of specific implementations with the goal that some of them might be compatible with quantum gravity. Refs.~\cite{Rudelius:2015xta,Montero:2015ofa} showed that the large effective axion constant that emerges only as an effective quantity in~\cite{Kim:2004rp}, can be made explicit by a different choice of basis, in which the potential is diagonal, and the large quantity emerges as the eigenvalue of the (non-diagonal) kinetic matrix. Furthermore, multi-field models of axion inflation have been studied~\cite{Rudelius:2015xta,Montero:2015ofa,Brown:2015iha,Bachlechner:2015qja,Hebecker:2015rya,Brown:2015lia,Heidenreich:2015wga,Kappl:2015esy,Choi:2015aem} in general under the \textit{weak gravity conjecture} (WGC) \cite{Arkani-Hamed:2006emk,Cheung:2014vva} which is expected to hold in a generic theory of quantum gravity. These works show that the minimal formulation of the alignment mechanism~\cite{Kim:2004rp} is not compatible with the WGC. However, additional instantons, might be present which, although subdominant in the potential, provide `additional interactions' that could make the alignment mechanism compatible with the WGC~\cite{Bachlechner:2015qja,Hebecker:2015rya,Brown:2015iha,Brown:2015lia}. This additional contributions might also generate subleading modulations to the potential, which can affect the CMB phenomenology~\cite{Kappl:2015esy,Choi:2015aem}. The embedding of the aligned mechanism in string theory has been further studied in Refs.~\cite{Long:2014dta,Gao:2014uha,Hebecker:2018fln,Palti:2015xra,Angus:2021jpr}.

Beyond the issue of its consistency with quantum gravity, it is also important to confront the predictions of the minimal model of aligned axion inflation~\cite{Kim:2004rp} with CMB measurements. Ref.~\cite{Peloso:2015dsa} showed that the inflationary trajectories in this model that end near minima of the potential~\footnote{For this reason, we denote these trajectories as ``stable'', as they terminate at a minimum of the potential as in typical single field models of slow-roll inflation.} are characterized by a tensor signal greater than that of natural inflation~\cite{Freese:1990rb}. These solutions are therefore also ruled out by data~\cite{Planck:2018jri}. On the other hand, it was found in~\cite{Peloso:2015dsa} that the model of~\cite{Kim:2004rp} also admits a second class of ``metastable''~\footnote{We use this term since, despite they terminate due to an instability, these trajectories extend for a number of e-folds much grater than those required for successful inflation.} inflationary trajectories, for which the end of inflation is caused by a tachyonic instability in the direction orthogonal to the inflationary trajectory. These solutions are characterized by a smaller tensor-to-scalar ratio than the ``stable'' ones, in agreement with observations. This arises because, for metastable trajectories, the majority of inflation occurs near a saddle point, where the potential is flatter. 

Contrary to the study of~\cite{Peloso:2015dsa}, which was mostly numerical, in this work we obtain  simple and compact analytic results for these trajectories and their associated CMB phenomenology. The latter can be summarized in eqs.~(\ref{ns-r-analytic}), that provide the scalar spectral tilt and the tensor-to-scalar ratio analytically, as a function of the model parameters, in a basis-independent way for the two axions. These simple expressions represent a significant advance over a numerical scan of a model that, in the original formulation of~\cite{Kim:2004rp} is characterized by $6$ parameters (two potential and four axion scales). The precise phenomenology (that, we stress, significantly differs from that of the more standard and more studied ``stable'' trajectories) is required to identify the viable parameter range, and to then precisely quantify the issue of compatibility of the concrete realizations of this mechanism with quantum gravity. 

This work is organized as follows: in Section \ref{sec:model} we introduce the model of aligned axion inflation~\cite{Kim:2004rp}, classifying the critical points of the potential. We introduce parameters which are invariant under a change of basis of the two axions. In Section \ref{sec:trajectory} we derive an analytic condition that characterizes the metastable trajectory, which we simplify in the aligned limit. In Section \ref{sec:traj-solution} we use this condition to determine the end of inflation and we construct the analytic one-field effective theory for inflation along the metastable trajectory, obtaining the result~(\ref{ns-r-analytic}) mentioned above. In Section~\ref{sec:numerical-CMB} we show that our analytic results reproduce very accurately those obtained from the exact numerical integration of the model. In performing the numerical integration, we also study, for some specific examples, the evolution of the fields after the inflationary trajectory becomes unstable. We find that, in some cases, the fields reach a stable trajectory connected to a minimum, so that inflation occurs in two separate stages for these solutions. In Section~\ref{sec:summary} we provide an overview of the results for a fixed degree of alignment and a fixed scalar spectral tilt (doing so, all the results for this model can be presented in terms of only two free parameters). In Section \ref{sec:transPlanck}, we discuss the trans-Planckian problem associated with the eigenvalues of the kinetic matrix in view of the new analytical results obtained above. Section~\ref{sec:concl} contains our conclusions. The work is ended by Appendix~\ref{app:eigensystem}, in which we present some intermediate results used in the main text for the analytic determination of the metastable trajectory.

\section{The model and the critical points of its potential}
\label{sec:model}

The model of aligned axion inflation~\cite{Kim:2004rp} is characterized by the lagrangian~\footnote{We work in natural units, with the line element $ds^2 = - d t^2 + a^2 \left( t \right) d \vec{x}^2$, where $a$ is the scale factor of the universe. In the following, $M_p$ and $H \equiv \frac{\ddot{a}}{a}$ denote, respectively, the reduced Planck mass, and the Hubble rate (where dot denotes derivative with respect to time $t$).}
\begin{equation}
{\cal L} = - \frac{1}{2} \left( \partial \theta \right)^2  - \frac{1}{2} \left( \partial \rho \right)^2 
- \sum_{i=1,2} \Lambda_i^4 \left[ 1 - \cos \left( \frac{\theta}{f_i} + \frac{\rho}{g_i} \right) \right] \,.
\label{knp}
\end{equation}
Without loss of generality, we sort the two potential terms by assuming that $\Lambda_1 \geq \Lambda_2$. Moreover, we preform the constant shifts~\footnote{As shown below, the shift sets a saddle point of the potential in the origin. This is the starting point of the metastable inflationary solutions that are the main interest of our work.}
\begin{equation}
\theta \equiv {\hat \theta} - \frac{g_2^{-1} \, \pi}{f_1^{-1} g_2^{-1} - g_1^{-1}  f_2^{-1}} \;\;,\;\; 
\rho \equiv {\hat \rho} + \frac{f_2^{-1} \, \pi}{f_1^{-1} g_2^{-1} - g_1^{-1}  f_2^{-1}} \;\;,
\end{equation}
in terms of which the lagrangian of the model reads
\begin{equation}
{\cal L} = - \frac{1}{2} \left( \partial {\hat \theta} \right)^2  - \frac{1}{2} \left( \partial {\hat \rho} \right)^2 
- \frac{\Lambda^4}{1+r_\Lambda} \left( 1 + r_\Lambda + \cos  {\cal A}_1 - r_\Lambda \, \cos  {\cal A}_2 \right) \;\;,\;\; {\cal A}_i \equiv f_i^{-1} \, {\hat \theta} + g_i^{-1} \, {\hat \rho} \,, 
\label{potential}
\end{equation}
where we also parametrized the potential scales through
\begin{equation}
\Lambda^4 \equiv \Lambda_1^4 + \Lambda_2^4 \;\;,\;\; r_\Lambda \equiv \frac{\Lambda_2^4}{\Lambda_1^4} \;. 
\label{Lambda-resca}
\end{equation}

The choice $\Lambda_1 \geq \Lambda_2$ rewrites $r_\Lambda \leq 1$. This choice does not completely fix the redundancy associated in the description of the model, that is still invariant under the internal rotation
\begin{eqnarray}
\left( \begin{array}{c}
{\hat \theta} \\ {\hat \rho} 
\end{array} \right) \;\to\;
\left( \begin{array}{c}
{\tilde \theta} \\ {\tilde \rho} 
\end{array} \right) \equiv R
\left( \begin{array}{c}
{\hat \theta} \\ {\hat \rho} 
\end{array} \right) \;\;,\;\; 
\left( \begin{array}{c}
f_i^{-1} \\ g_i^{-1} 
\end{array} \right) \;\to\;
\left( \begin{array}{c}
{\tilde f}_i^{-1} \\ {\tilde g}_i^{-1} 
\end{array} \right) \equiv R
\left( \begin{array}{c}
f_i^{-1} \\ g_i^{-1} 
\end{array} \right) \;\;,\;\;  i = 1,\,2 \;,
\label{rotation}
\end{eqnarray}
where $R$ is a $2 \times 2$ global rotation matrix. Physical results need to be invariant under this rotation, and the goal of this study is to express the final phenomenological results in terms of the invariant combinations~\footnote{In practice, we introduce the two-dimensional internal vectors $\left\{ f_i^{-1} ,\, g_i^{-1} \right\}$, and the combinations in (\ref{NC}) are their squared norms and their vector product.}
\begin{equation}
n_1 \equiv f_1^{-2} + g_1^{-2} \;\;,\;\; n_2 \equiv f_2^{-2} + g_2^{-2} \;\;,\;\; {\cal C} \equiv f_2^{-1} g_1^{-1} - f_1^{-1} g_2^{-1} \;.
\label{NC}
\end{equation}
The quantity ${\cal C}$ vanishes in the aligned limit $f_1/f_2 = g_1/g_2$, and in the following we will parametrize the degree of alignment with the dimensionless combination
\begin{equation}
\gamma \equiv \frac{2 \, {\cal C}}{n_1+n_2} \;,
\label{gamma}
\end{equation}
which satisfies $\vert \gamma \vert \ll 1$ in the limit of nearly aligned potential and which replaces the combination $\alpha$ introduced in~\cite{Peloso:2015dsa}, which is not invariant under~(\ref{rotation}).~\footnote{In the following, for brevity, we present expressions expanded in the near aligned limit as expansions in ${\cal C}$. Since ${\cal C}$ is a dimensionful parameter, the following expansions should be more properly understood as expansions in $\gamma$, in the limit of small $\left\vert \gamma \right\vert$.} 

\begin{figure}
\centering
\includegraphics[scale=0.6]{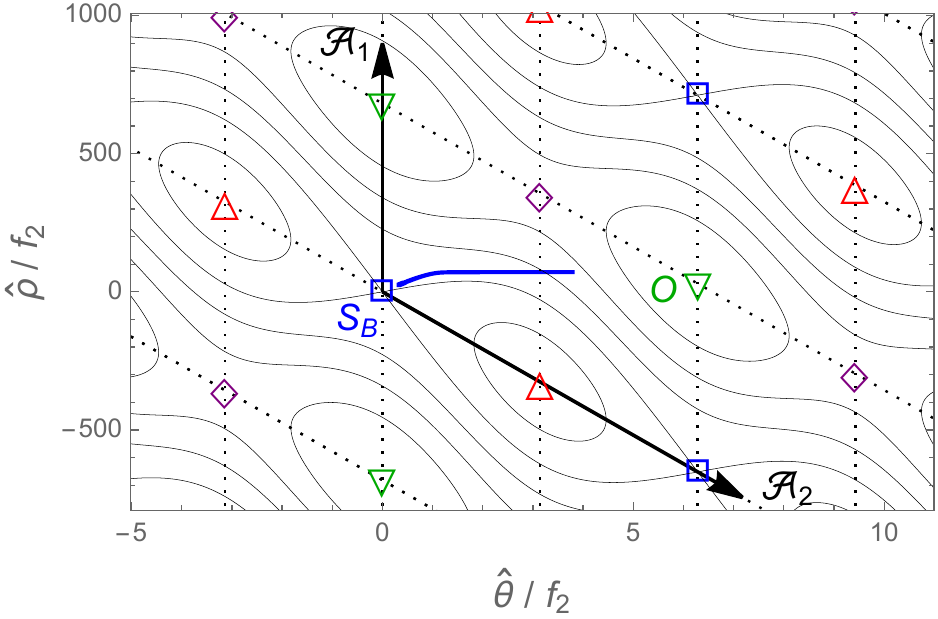}
\caption{Contour plot of the potential for a specific example with alignment $\gamma = 0.0075$ in eq.~(\ref{gamma}) (note the different ranges shown in the two axes). The figure marks maxima (upwards red triangles), minima (downwards green triangles), saddle points ${\cal S}_A$ (purple diamonds), saddle points ${\cal S}_B$ (blue squares), and symmetry domains (dotted lines) of the potential. The blue line shows the last $60$ e-folds of a metastable inflationary trajectory starting from a neighbourhood of the saddle point marked as ${\cal S}_B$ (the post-inflationary trajectory, not shown in the figure, is characterized by damped oscillations about the minimum marked as ${\cal O}$). This trajectory leads to the spectral tilt $n_s = 0.965$ and $r = 2 \times 10^{-3}$ at $60$ e-folds before the end of inflation.
}
\label{fig:plotV}
\end{figure}

Due to the symmetry under a shift of its arguments ${\cal A}_i \to {\cal A}_i + 2 \pi$, and to the even parity under the change of sign ${\cal A}_i \to - {\cal A}_i$, it is enough to consider the potential in (\ref{potential}) in the $0 \leq {\cal A}_1 \leq \pi \;,\; 0 \leq {\cal A}_2 \leq \pi$ domain. It is immediate to see that, apart from the degenerate case of a perfect alignment, all the critical points of the potential occur for $\sin {\cal A}_1 = \sin {\cal A}_2 = 0$. In the domain, there are four such points that, using the notation of~\cite{Peloso:2015dsa}, we indicate as ${\cal O}$, corresponding to a minimum of the potential, as ${\cal S}_{A,B}$, corresponding to two saddle points, and as ${\cal M}$, corresponding to a maximum. We actually indicate with these letters all the critical points of the potential, also beyond this single domain. Table~\ref{tab:critical} lists the values of the two arguments of the cosines, of the potential, and of the eigenvalues of the Hessian (the second derivative of the potential with respect to the two fields) corresponding to each critical point. Clearly, all these values are invariant under the internal rotation~(\ref{rotation}), and we could therefore express them in terms of the invariant parameters (\ref{NC}). Figure~\ref{fig:plotV} shows an example of aligned inflationary potential ($\gamma = 0.0075$) leading to an inflationary trajectory with an acceptable CMB  phenomenology (see Section~\ref{sec:numerical-CMB} for details).

\renewcommand\arraystretch{2} 
\begin{table}
\begin{tabular}{|c|c|c|c|}
  \hline
  Critical point & $\left\{ {\cal A}_1 ,\, {\cal A}_2 \right\} / \pi$ & $V / \frac{2 \Lambda^4}{1+r_\Lambda}$ & $V'' / \frac{\Lambda^4}{2 \left( 1+r_\Lambda \right)}$ \\
  \hline\hline
  Minimum ${\cal O}$ & $\left\{ {\rm odd} ,\, {\rm even} \right\}$ & $0$ & $n_1 + r_\Lambda \, n_2 \pm \sqrt{ \left( n_1 + r_\Lambda \, n_2 \right)^2 - 4 r_\Lambda {\cal C}^2}$\\
  \hline
  Saddle ${\cal S}_A$ & $\left\{ {\rm odd} ,\, {\rm odd} \right\}$ & $r_\Lambda$ & $n_1 - r_\Lambda \, n_2 \pm \sqrt{ \left( n_1 - r_\Lambda \, n_2 \right)^2 + 4 r_\Lambda {\cal C}^2}$ \\
  \hline
  Saddle ${\cal S}_B$ & $\left\{ {\rm even} ,\, {\rm even} \right\}$ & $1$ & $-$ expressions for ${\cal S}_A$ \\
  \hline
  Maximum ${\cal M}$ & $\left\{ {\rm even} ,\, {\rm odd} \right\}$ & $ 1 + r_\Lambda$ & $-$ expressions for ${\cal O}$ \\
  \hline
\end{tabular}
\caption{Critical points of the potential in (\ref{potential}). The second, third, and fourth column, indicate, respectively, the value of the two arguments of the cosines ($\pi$ times an even or an odd integer), of the potential, and of the two eigenvalues of the Hessian, all evaluated at that critical point listed in the first column. 
}
\label{tab:critical}
\end{table}

As studied in~\cite{Peloso:2015dsa}, the potential in~(\ref{potential}) admits a first class of inflationary trajectories that end in a neighbourhood of a minimum ${\cal O}$. Their behaviour is analogous to that of the single axion natural inflation model~\cite{Freese:1990rb}, and, analogously to what happens for this model, they are also ruled out by the CMB data~\cite{Peloso:2015dsa}. The potential in~(\ref{potential}) also possesses a more phenomenologically interesting second class of solutions, that terminate not because they reach the minimum of the potential, but because the direction orthogonal to the inflationary one becomes unstable. 

For trajectories of the first class, the potential cannot be made arbitrarily flat (namely, arbitrarily small values of the slow roll quantities $\left( V'/V \right)^2$ and $\left\vert V'' / V \right\vert$) at observable scales by arbitrarily increasing the axion scales (or the degree of alignment). To see this, consider the single axion potential with the ratio $\frac{\phi}{f}$ in the argument of the cosine, and denote by $\phi_{\rm CMB}$ the value assumed by the inflaton $N_{\rm CMB}$ e-folds before the end of inflation, when the CMB modes were produced. For small $f$, the quantity $\phi_{\rm CMB}$ approaches the maximum of the potential $\phi = \pi f$. As $f$ is increased, the ratio $\phi_{\rm CMB}/f$ becomes progressively smaller, until the inflaton only spans the very bottom of the potential, where the potential can be well approximated by its quadratic term. In this regime, $\phi_{\rm CMB}$ becomes independent of $f$, and this scale also drops from the $V'/V$ and $V''/V$ ratios. 

Loosely speaking, it is true that the potential near a maximum is arbitrarily flat in the limit of arbitrary large axion scales. However, the region near the maximum becomes irrelevant for CMB phenomenology, as an increasingly large number of e-folds needs to pass in order for the inflaton to roll from a neighborhood of the maximum to  $\phi_{\rm CMB}$. This is not necessarily the case for solutions of the second class, that can spend most of inflation in a neighbourhood of the saddle point~\cite{Peloso:2015dsa}. Increasing the axion scale (or the degree of alignment), flattens the potential next to the saddle point, but does not automatically increase too much the number of e-folds needed to reach the end of inflation, since this does not occur due to a smooth increase of the slow roll parameters along the inflationary trajectory, but due to a more sudden instability in the orthogonal direction.

We denote this second class of solutions that start near a saddle point and terminate due to the instability of the orthogonal direction as `metastable inflationary trajectories'.~\footnote{In fact in this work we also show the presence of a third class of inflationary solutions, not noted in~\cite{Peloso:2015dsa}, for which the post inflationary evolution after the metastable trajectory does not immediately lead to a minimum, but to a stable inflationary valley connected to a minimum. Namely, these solutions are characterized by two separate inflationary stages.} We study these solutions analytically in the next section. Here we discuss one necessary condition for their existence, namely the request that one squared eigenmass at the saddle point is light and negative, while the other one is heavy and positive.~\footnote{The squared eigenmasses are the eigenvalues of the Hessian of the potential. Their exact expression is reported in Table~\ref{tab:critical}. In this discussion, ``heavy'' and `light' mean, respectively, an eigenmass that is of ${\rm O} \left( {\cal C}^0 \right)$ and of ${\rm O} \left( {\cal C}^2 \right)$ in the alignment.} The inflationary trajectory then proceeds along the eigenvector corresponding to the light mass, which is a direction that would be flat in the limit of perfect alignment. The stability of this direction near the saddle point is due to the fact that the other squared eigenmass is large and positive. 

We note from Table~\ref{tab:critical} that, for $n_1 > r_\Lambda \, n_2$, the squared eigenmasses at the saddle point ${\cal S}_A$ are one heavy and positive and one light and negative
\begin{equation}
n_1 > r_\Lambda \, n_2 \;\; \Rightarrow \;\; m_{\rm heavy}^2 \Big\vert_{{\cal S}_A} = + \frac{\left( n_1 - r_\Lambda \, n_2 \right) \Lambda^4}{1+r_\Lambda} + {\rm O } \left( {\cal C}^2 \right) \;\;,\;\; m_{\rm light}^2 \Big\vert_{{\cal S}_A} = - \frac{r_\Lambda \, \Lambda^4 \, {\cal C}^2}{\left( 1 + r_\Lambda \right) \left( n_1 - r_\Lambda \, n_2 \right)} + {\rm O } \left( {\cal C}^4 \right) \;,
\label{masses-SA}
\end{equation}
while the squared eigenmasses at the saddle point ${\cal S}_B$ are one light and positive and one heavy and negative (which does not correspond to a stable inflationary trajectory making use of the alignment mechanism of~\cite{Kim:2004rp}). The situation is reversed for $r_\Lambda \, n_2 > n_1$, in which the necessary conditions for a trajectory in the limit of near alignment are met at the other saddle point~\footnote{We disregard the special case $n_1 = r_\Lambda \, n_2$ for which no orthogonal heavy direction to the inflationary one is present at the saddle point.} 
\begin{equation}
r_\Lambda \, n_2 > n_1 \;\; \Rightarrow \;\; m_{\rm heavy}^2 \Big\vert_{{\cal S}_B} = + \frac{\left( r_\Lambda \, n_2 - N_1 \right) \Lambda^4}{1+r_\Lambda} + {\rm O } \left( {\cal C}^2 \right) \;\;,\;\; m_{\rm light}^2 \Big\vert_{{\cal S}_B} = - \frac{r_\Lambda \, \Lambda^4 \, {\cal C}^2}{\left( 1 + r_\Lambda \right) \left( r_\Lambda \, n_2 - n_1 \right)} + {\rm O } \left( {\cal C}^4 \right) \;.
\label{masses-SB}
\end{equation}

In the next section we determine analytically the inflationary trajectories  that start along the direction corresponding to the light negative squared eigenmass at the saddle point ${\cal S}_A$ (respectively, ${\cal S}_B$) in the case of $n_1 > r_\Lambda \, n_2$ (respectively, $r_\Lambda \, n_2 > n_1$). We investigate whether they are metastable, namely whether the orthogonal direction becomes tachyonic at any point during this trajectory.

\section{Analytic determination of the metastable inflationary trajectory}
\label{sec:trajectory}

In this section, we determine and study the analytic conditions for the existence and the stability of the inflationary trajectories streaming off the two saddle points. Ref.~\cite{Peloso:2015dsa} studied the model in the limit of strong alignment, and observed numerically that only the solution steaming off ${\cal S}_B$ is metastable, in the parameter range $\frac{1}{r_f^4} < r_\Lambda < \frac{1}{r_f^2}$, where $r_f \equiv \sqrt{f_1/f_2} \simeq \sqrt{g_1/g_2}$ in the limit of large alignment. We want to understand this observation and prove the condition analytically.

In the limit of large alignment, $m_{\rm heavy}^2 \gg \left\vert m_{\rm light}^2 \right\vert$ along the trajectory. Therefore, inflation proceeds along the light direction, without exciting the heavy direction. In the limit of slow roll, the inflaton evolves along the gradient of the potential, $\dot{\vec{\phi}} \simeq - \frac{1}{3 H} \vec{\nabla} V$, where $\dot{\vec{\phi}}$ and $\vec{\nabla} V$ are, respectively, the inflaton velocity and the gradient of the potential in the $\left\{ {\hat \theta} ,\, {\hat \rho} \right\}$ plane. We denote by $\vec{v}_{\rm heavy}$ and $\vec{v}_{\rm light}$ the eigenvectors corresponding to, respectively, the heavy and light squared eigenmasses. Since the Hessian is hermitian, and the two eigenmasses are different, these two eigenvectors are orthogonal. Therefore, the statement that the trajectory occurs along the light direction can be mathematically expressed as
\begin{equation}
\vec{v}_{\rm heavy} \cdot \vec{\nabla} V \;\Big\vert_{\rm trajectory} = 0 \;. 
\label{def-traj}
\end{equation}
This condition, which is approximately true in the limits of slow roll and near alignment, determines the trajectory. The exact expression corresponding to (\ref{def-traj}) is eq.~(\ref{traj}) of Appendix~\ref{app:eigensystem}. We show there that, in the limit of strong alignment, this expression simplifies to
\begin{equation}
{\rm trajectory}:\;\; 
-\sin {\cal A}_1 + r_\Lambda \, \sqrt{\frac{n_2}{n_1}} \, \sin {\cal A}_2 = 0 \;,
\label{trajectory}
\end{equation}
which, as expected, is invariant under the rotation (\ref{rotation}). The trajectory is stable as long as $m_{\rm heavy}^2 \geq 0$, where the equality determines the point at which the orthogonal direction becomes unstable. Eq.~(\ref{eigen-mass}) of Appendix~\ref{app:eigensystem} (with the plus sign) is the exact expression for the heavy eigenmass. We show there that, in the limit of strong alignment, the condition $m_{\rm heavy}^2 \geq 0$ gives~\footnote{More accurately, the trajectory becomes unstable when the orthogonal squared mass becomes small (even if it is still positive). The simplified condition (\ref{stability}) amounts in associating the start of the instability to the moment in which the ${\rm O } \left( {\cal C}^0 \right)$ contribution to the orthogonal mass vanishes. As we show numerically, this provides a very good estimate for the end of the inflationary trajectory. Due to the strong hierarchy between the ${\rm O } \left( {\cal C}^0 \right)$ and the ${\rm O } \left( {\cal C}^2 \right)$ contributions, in the text, for brevity, we will simply refer to the vanishing of the ${\rm O } \left( {\cal C}^2 \right)$ contribution as to the ``vanishing of the heavy eigenmass''.}
\begin{equation}
{\rm stablity}:\;\; 
-\cos {\cal A}_1 + r_\Lambda \, \frac{n_2}{n_1} \, \cos {\cal A}_2 \geq 0 \;,
\label{stability}
\end{equation}
which, obviously, is also invariant under (\ref{rotation}). Inflation ends when the last two expressions vanish
\begin{equation}
{\rm end \; of \; inflation}: \;\; 
\left\{ \begin{array}{l}
-\sin {\cal A}_1 + r_\Lambda \, \sqrt{\frac{n_2}{n_1}} \, \sin {\cal A}_2 = 0 \;, \\ 
-\cos {\cal A}_1  + r_\Lambda \, \frac{n_2}{n_1} \, \cos {\cal A}_2 = 0 \;. 
\end{array} \right.
\label{end-1}
\end{equation}
Let us study when this system can be solved. From the two relations, it is immediate to write
\begin{equation}
\sin^2 {\cal A}_1 + \cos^2 {\cal A}_1 = 
\frac{r_\Lambda^2 \, n_2}{n_1} \sin^2 {\cal A}_2 +  
\frac{r_\Lambda^2 \, n_2^2}{n_1^2} \cos^2 {\cal A}_2 = 1 \;. 
\end{equation}

From the last equal sign, we obtain
\begin{equation}
\cos^2 {\cal A}_2 =  \frac{n_1 \left( r_\Lambda^2 \, n_2 - n_1 \right)}{r_\Lambda^2 \, n_2 \left( n_1 - n_2 \right)} \;\;,\;\; \sin^2 {\cal A}_2 = \frac{n_1^2 - r_\Lambda^2 \, n_2^2}{r_\Lambda^2 \, n_2 \left( n_1 - n_2 \right)} \;. 
\label{cA2-sA2}
\end{equation}
We note that the two right hand sides (r.h.s.) indeed add up to $1$; therefore the system has a solution if and only if there are parameters for which both r.h.s. are positive. We now show that, due to the $r_\Lambda \leq 1$ choice done at the beginning of our analysis, this is not possible for $n_1 > r_\Lambda \, n_2$. In this case, the numerator of the second expression is positive, forcing also the denominator to be positive, $n_1 > n_2$. We then see that also the first numerator needs to be positive, $r_\Lambda^2 \, n_2 > n_1$. The last two conditions sum up to $n_2 < n_1 < r_\Lambda^2 \, n_2$. This is impossible, due to the fact that $r_\Lambda \leq 1$. 

In the complementary case of $r_\Lambda n_2 > n_1$ the second numerator is negative, forcing the denominator to be also negative, $n_2 > n_1$, and then also the first numerator to be negative, $n_1 > r_\Lambda^2 n_2$. These conditions can be summarized as
\begin{equation}
r_\Lambda^2 < \frac{n_1}{n_2} < r_\Lambda \;, 
\label{cond-metastble}
\end{equation}
which reproduces the condition observed in \cite{Peloso:2015dsa} and discussed at the beginning of this section.  

We have therefore proven that a metastable solution exists streaming off the saddle point ${\cal S}_B$ provided that the condition (\ref{cond-metastble}) is met, in the limit of large alignment. No metastable solution streams off ${\cal S}_A$.~\footnote{As already mentioned, the choice $r_\Lambda \leq 1$ does not restrict the generality of our study, as the model~(\ref{knp}) is invariant under the simultaneous exchange of $r_\Lambda \leftrightarrow \frac{1}{r_\Lambda} ,\, f_1 \leftrightarrow f_2 ,\, g_1 \leftrightarrow g_2$. The present discussion shows that, for $r_\Lambda > 1$, the saddle point from which the metastable inflationary trajectory streams off is the one denoted by ${\cal S}_A$ in our notation.} This proves analytically what was observed numerically in~\cite{Peloso:2015dsa}.  

\section{Effective $1$d theory along the trajectory} 
\label{sec:traj-solution}

In the previous section, we have proven that, in the parameter range (\ref{cond-metastble}), the model admits the metastable trajectory streaming off the saddle point ${\cal S}_B$. It is straightforward to compute numerically the two-field evolution along the metastable inflationary trajectory~\cite{Peloso:2015dsa}. Our goal is to obtain simple approximate analytical expressions for the trajectory and the associated phenomenology.

Most of the inflationary trajectory occurs next to the saddle point, where it is sufficient to retain the expansion of the potential to quadratic order in the displacement from this point. The expansion is immediately obtained from the values of the potential and the light eigenvalue reported in Table~\ref{tab:critical}, leading to the quadratic effective $1$ field potential 
\begin{equation}
V_{\rm eff} \left( \varphi \right) = \frac{2 \, \Lambda^4}{1+r_\Lambda} \left\{ 1 - \left[ \frac{r_\Lambda \, {\cal C}^2}{4 \left( r_\Lambda \, n_2 - n_1 \right)} + {\rm O } \left( {\cal C}^4 \right) \right] \varphi^2+ {\rm O } \left( \varphi^4 \right) \right\} \;,
\label{V-eff}
\end{equation}
where $\varphi$ is the canonically normalized field parametrizing the inflationary trajectory and that evaluates to $\varphi = 0$ at the saddle point. The light quadratic mass term has been expanded in the near alignment limit, cf. eq.~(\ref{masses-SB}). We note that the effective potential (\ref{V-eff}) is manifestly invariant under the rotation~(\ref{rotation}). 

As we verify below, most of the inflationary trajectory occurs near the saddle point, where the quadratic potential~(\ref{V-eff}) is sufficiently adequate. The slow roll parameters associated to this potential are
\begin{eqnarray}
\epsilon &\equiv& \frac{M_p^2}{2} \, \left( \frac{1}{V} \, \frac{d V}{d \varphi} \right)^2 = \left[ \frac{M_p^2 \, r_\Lambda^2 \, {\cal C}^4}{8 \left( r_\Lambda \, n_2 - n_1 \right)^2} + {\rm O } \left( {\cal C}^6 \right) \right] \varphi^2 + {\rm O } \left(  \varphi^4 \right) \;, \nonumber\\
\eta &\equiv& \frac{M_p^2}{V} \, \frac{d^2 V}{d \varphi^2} = \left[ - \frac{M_p^2 \, r_\Lambda \, {\cal C}^2}{2 \left( r_\Lambda \, n_2 - n_1 \right)} + {\rm O } \left( {\cal C}^4 \right) \right] + {\rm O } \left(  \varphi^2 \right) \;.
\label{epsilon-eta}
\end{eqnarray}
These quantities allow us to evaluate the spectral scalar tilt and the tensor-to-scalar ratio in slow roll approximation. We find, respectively,
\begin{eqnarray}
n_s - 1 &\simeq& - 6 \epsilon + 2 \eta = \left[ - \frac{M_p^2 \, r_\Lambda \, {\cal C}^2}{r_\Lambda \, n_2 - n_1} + {\rm O } \left( {\cal C}^4 \right) \right] + {\rm O } \left(  \varphi_N^2 \right) \;, \nonumber\\
r &\simeq& 16 \epsilon = \left[ \frac{2 M_p^2 \, r_\Lambda^2 \, {\cal C}^4}{\left( r_\Lambda \, n_2 - n_1 \right)^2} + {\rm O } \left( {\cal C}^6 \right) \right] \varphi_N^2 + {\rm O } \left(  \varphi_N^4 \right) \;.
\label{ns-r}
\end{eqnarray}
It is worth noting that the tensor-to-scalar ratio is more suppressed than the spectral tilt, both by the smallness of ${\cal C}$ and of $\varphi_N$. This is the reason why, unlike in the original natural inflation model~\cite{Freese:1990rb}, the phenomenology associated with these trajectories is still consistent with the data. We also note that, to leading order, the spectral tilt receives a contribution only from $\eta$, and it is constant (namely, scale invariant). On the other hand, the tensor-to-scalar ratio is scale dependent, as stressed by the suffix $N$ that we have added to the inflaton, meaning that we evaluate it $N$ e-folds before the end of inflation, when the corresponding scale left the horizon. 

The $N$ dependence of $\phi$ during inflation can be obtained from the slow roll equation of motion of the inflaton field, $3 H \dot{\varphi} \simeq - \frac{d V}{d \varphi}$ and from the relation $d N = - H \, d t$. Using also the slow roll relation $H^2 \simeq \frac{V}{3 M_p^2}$, one arrives to
\begin{equation}
\frac{1}{M_p} \frac{d \varphi}{d N} \simeq - \sqrt{2 \epsilon} \;\; \Rightarrow\;\; \frac{1}{M_p} \int_{\varphi_{N_1}}^{\varphi_{N_2}} \frac{d \varphi}{\sqrt{2 \epsilon}} \simeq - \left( N_2 - N_1 \right)\;,
\end{equation}

In the quadratic potential, from the first eq. of (\ref{epsilon-eta}) at leading order, we obtain
\begin{equation}
\frac{1}{\sqrt{2 \epsilon}} \simeq \frac{2 M_p}{\left\vert n_s - 1 \right\vert} \, \frac{1}{\varphi} \;\; \Rightarrow \;\; \varphi_{N_1} \simeq \varphi_{N_2} \, {\rm e}^{- \frac{\left\vert n_s - 1 \right\vert}{2} \left( N_1 - N_2 \right)} \;. 
\label{phi-N}
\end{equation}
In this relation, `2' refers to a time greater than `1', namely closer to the end of inflation. Therefore $N_2 < N_1$. This relation correctly indicates that $\varphi_{N_1} < \varphi_{N_2}$, namely that the inflaton increases during inflation. 

The quadratic expansion leading to (\ref{V-eff}) is not perfectly adequate all throughout inflation, as the eigenmasses of the quadratic theory are constant, and so this expansion cannot account for the end of inflation, that occurs when the heavy eigenmass vanishes. Still, most of the inflationary trajectory occurs close to the saddle point (where the inflaton moves more slowly) and, as we verify numerically in the next section, $\varphi_{N_{\rm CMB}}$ is in the region where the expansion is valid, and the relations (\ref{ns-r}) are accurate. On the other hand, this implies that, strictly speaking, the relation~(\ref{phi-N}) cannot be valid at the end of inflation. Still, due to the fact that most of inflation occurs in the regime in which the quadratic approximation is accurate and that the derivatives of the full potential do not change abruptly, we expect that using it all throughout inflation leads to a reasonably good estimate for $\varphi_{N_{\rm CMB}}$ and, ultimately, for the tensor-to-scalar ratio (as we explicitly verify in the next two sections). Specifically, we estimate
\begin{equation}
\varphi_N \simeq \varphi_{\rm end} \, {\rm e}^{- \frac{\left\vert n_s - 1 \right\vert}{2} N}  
\;\;\; \Rightarrow \;\;\; 
r \simeq 2 \left\vert n_s -1 \right\vert^2 \, \frac{\varphi_{\rm end}^2}{M_p^2} \, {\rm e}^{- \left\vert n_s - 1 \right\vert N} \;.
\label{phiN}
\end{equation} 

\subsection{Analytic expressions for $\varphi_{\rm end}$, $n_s-1$ and $r$}

To make use of the estimate (\ref{phiN}) we need an explicit expression for the value $\varphi_{\rm end}$ of the normalized inflaton field at the end of inflation. We achieve this in two steps: we first solve the system (\ref{end-1}) and we then explicitly relate the values of the two fields to that of $\varphi$. To perform the first step, we obtain ${\cal A}_2$ from the first of~(\ref{cA2-sA2}), and then  ${\cal A}_1$ from the first of~(\ref{end-1}). We then recall the definitions of ${\cal A}_{1,2}$ in eq.~(\ref{potential}) to write 
\begin{eqnarray}
{\hat \theta}_{\rm end} &=& \frac{1}{g_1 \, {\cal C}} \, \arccos \left( \frac{1}{r_\Lambda} \, \sqrt{\frac{n_1 \left( n_1 - r_\Lambda^2 \, n_2 \right)}{n_2 \left( n_2 - n_1 \right)}} \right) - \frac{1}{g_2 \, {\cal C}} \, \arcsin \left( \sqrt{\frac{r_\Lambda^2 n_2^2 - n_1^2}{n_1 \left( n_2 - n_1 \right)}} \right) \;, \nonumber\\
{\hat \rho}_{\rm end} &=& - \frac{1}{f_1 \, {\cal C}} \, \arccos \left( \frac{1}{r_\Lambda} \, \sqrt{\frac{n_1 \left( n_1 - r_\Lambda^2 \, n_2 \right)}{n_2 \left( n_2 - n_1 \right)}} \right) + \frac{1}{f_2 \, {\cal C}} \, \arcsin \left( \sqrt{\frac{r_\Lambda^2 n_2^2 - n_1^2}{n_1 \left( n_2 - n_1 \right)}} \right) \;,\;\;\; r_\Lambda^2 < \frac{n_1}{n_2} < r_\Lambda \;, \nonumber\\ 
\label{sol-th-rh}
\end{eqnarray}
where we also recalled the conditions (\ref{cond-metastble}) under which this solution exists. 

In writing this solution, we could equivalently have expressed ${\cal A}_1$ through an arccosine and / or ${\cal A}_2$ through an arcsine. Our choice has been motivated by the fact that the two arguments of the arc functions in (\ref{sol-th-rh}) are smaller than $1$ in the range of their validity. More precisely, the argument of the arccosine (respectively, arcsine) retained in (\ref{sol-th-rh}) grows from $0$ to $1$ (respectively, decreases from $1$ to $0$) as the ratio $\frac{n_1}{n_2}$ increases from $r_\Lambda^2$ to $r_\Lambda$. For $r_\Lambda^2 \ll \frac{n_1}{n_2} \ll r_\Lambda$, both arguments are much smaller than $1$. 

As the second step to obtain $\varphi_{\rm end}$, we need to relate $\varphi$ to the fields ${\hat \theta}$ and ${\hat \rho}$. A formal procedure to do this is the following: obtain the relation ${\hat \rho}_{\rm traj} \left( {\hat \theta} \right)$ from the trajectory condition~(\ref{traj}); then, recalling that $\varphi$ is canonically normalized, express $d \varphi = \sqrt{d {\hat \theta}^2  + d {\hat \rho}^2} = \sqrt{1 + \left( \frac{d {\hat \rho}_{\rm traj}}{d {\hat \theta}} \right)^2} \, d {\hat \theta}$; integrate this expression, so to obtain $\varphi$ as a function of ${\hat \theta}$ along the trajectory; finally evaluate this relation at ${\hat \theta}_{\rm end}$. This can be done analytically only within some approximation scheme. We chose the approximation of small departure from the saddle point, which is the approximation under which (\ref{phiN}) is valid. Then, the linear relation between the fields can be more immediately obtained by the fact that, next to the saddle point, the evolution proceeds along the light eigenvector $\vec{v}_-$ given in eq.~(\ref{eigen-vector}). Namely, denoting by $i=1,2$ the component of this vector along the ${\hat \theta}$ and ${\hat \rho}$ directions, we have the linear relations (valid in the proximity of the saddle point), ${\hat \theta} = \vec{v}_{-,1} \, \varphi$ and ${\hat \rho} = \vec{v}_{-,2} \, \varphi$. Evaluating $\vec{v}_{-,i}$ at ${\cal S}_B$, and expanding the two components in the limit of strong alignment, we obtain 
\begin{eqnarray}
{\hat \theta} &=& \left[ - \frac{1}{g_1 \, \sqrt{n_1}} + \frac{\sqrt{n_2} \, r_\Lambda \, {\cal C}}{f_1 \, n_1 \left( r_\Lambda \, n_2 - n_1 \right)} + {\rm O } \left( {\cal C}^2 \right) \right] \varphi + {\rm O } \left( \varphi^3 \right) \;, \nonumber\\
{\hat \rho} &=& \left[  \frac{1}{f_1 \, \sqrt{n_1}} + \frac{\sqrt{n_2} \, r_\Lambda \, {\cal C}}{g_1 \, n_1 \left( r_\Lambda \, n_2 - n_1 \right)} + {\rm O } \left( {\cal C}^2 \right) \right] \varphi + {\rm O } \left( \varphi^3 \right) \;.
\end{eqnarray}

We can obtain $\varphi_{\rm end}$ by inverting the first expression and evaluating it at the value ${\hat \theta}_{\rm end}$ given in (\ref{sol-th-rh}), or, alternatively, by inverting the second expression and evaluating it at the value ${\hat \rho}_{\rm end}$ given in (\ref{sol-th-rh}). The two resulting expressions for $\phi_{\rm end}$ coincide to leading order in the alignment: 
\begin{equation}
\varphi_{\rm end} = \frac{1}{\cal C} \left[ \sqrt{n_2} \, \arcsin \left( \sqrt{\frac{r_\Lambda^2 n_2^2 - n_1^2}{n_1 \left( n_2 - n_1 \right)}} \right) - \sqrt{n_1} \, \, \arccos \left( \frac{1}{r_\Lambda} \, \sqrt{\frac{n_1 \left( n_1 - r_\Lambda^2 \, n_2 \right)}{n_2 \left( n_2 - n_1 \right)}} \right) \right] + {\rm O } \left( {\cal C}^0 \right) \;\;,\;\; r_\Lambda^2 < \frac{n_1}{n_2} < r_\Lambda \;. 
\label{phiend1}
\end{equation}
This result is manifestly invariant under the internal rotation~(\ref{rotation}). 

Inserting this into the second of~(\ref{phiN}), and recalling the first of~(\ref{ns-r}), we obtain our final analytic approximate expressions for the CMB phenomenology associated to the metastable trajectories, valid in the limit of strong alignment:
\begin{eqnarray}
{\rm For} && r_\Lambda^2 \leq \frac{n_1}{n_2} \leq r_\Lambda \,: \nonumber\\\nonumber\\
&& n_s - 1 \simeq - \frac{M_p^2 \, r_\Lambda \, {\cal C}^2}{r_\Lambda \, n_2 - n_1} \;,\nonumber\\
&& r \simeq  \frac{2 \left\vert n_s -1 \right\vert^2}{{\cal C}^2 \, M_p^2} \, \left[ \sqrt{n_2} \, \arcsin \left( \sqrt{\frac{r_\Lambda^2 n_2^2 - n_1^2}{n_1 \left( n_2 - n_1 \right)}} \right) - \sqrt{n_1} \, \, \arccos \left( \frac{1}{r_\Lambda} \, \sqrt{\frac{n_1 \left( n_1 - r_\Lambda^2 \, n_2 \right)}{n_2 \left( n_2 - n_1 \right)}} \right) \right]^2 \, {\rm e}^{- \left\vert n_s - 1 \right\vert N} \;.\nonumber\\
\label{ns-r-analytic}
\end{eqnarray}

As discussed above, for $r_\Lambda^2 \ll \frac{n_1}{n_2} \ll r_\Lambda$, the arguments of the arccosine and arcsine are much smaller than $1$, and we obtain the extremely simplified relations
\begin{equation}
r_\Lambda^2 \ll \frac{n_1}{n_2} \ll r_\Lambda \;\; \Rightarrow \;\; 
\varphi_{\rm end} \simeq  \frac{r_\Lambda \, n_2}{\sqrt{n_1} \, {\cal C}} \;\;\;,\;\;\; 
r \simeq \frac{2 \, r_\Lambda^2 \, n_2^2 \, \left\vert n_s -1 \right\vert^2}{n_1 \, {\cal C}^2 \, M_p^2} \, 
{\rm e}^{- \left\vert n_s - 1 \right\vert N}  \;.
\end{equation}

\section{Comparison with the numerical results}
\label{sec:numerical-CMB}

In this section, we provide the set-up to study the $2-$field evolution numerically and compare against the analytic results (\ref{phiend1}) and (\ref{ns-r-analytic}). As already done in Figure~\ref{fig:plotV}, we fix the freedom of internal rotations~(\ref{rotation}) by setting $g_2^{-1} = 0$. We then introduce the following dimensionless fields and parameters
\begin{equation}
{\tilde \theta} \equiv \sqrt{n_2} \, {\hat \theta} \;,\; 
{\tilde \rho} \equiv \sqrt{n_2} \, {\hat \rho} \;,\; 
{\tilde n}_2 \equiv M_p^2 \, n_2 \;,\; \alpha \equiv \frac{\sqrt{n_1 \, n_2 - {\cal C}^2}}{n_2} \;,\; 
\beta \equiv \frac{\cal C}{n_2} \;,
\end{equation}
in terms of which the potential reads
\begin{equation}
V = \frac{\Lambda^4}{2 \left( 1+r_\Lambda \right)} \left[  1 - \cos \left( \alpha \, {\tilde \theta} + \beta \, {\tilde \rho} \right) + r_\Lambda  \left( 1 - \cos \, {\tilde \theta} \right) \right] \;. 
\end{equation}
We use the number of e-folds ${\hat N}$ as time variable in the numerical evolution, with the convention that it grows during inflation, $d {\hat N} = H \, d t$ (hence ${\hat N}$ is minus the number of e-folds $N$ appearing in the analytic relations of the previous sections). Then, starting from slow roll initial conditions, we integrate the evolution equations
\begin{eqnarray}
&& {\tilde \theta}'' + \left( 3 - \frac{{\tilde \theta}^{'2} + {\tilde \rho}^{'2}}{2 \, {\tilde n}_2} \right) {\tilde \theta}'  + \frac{3 {\tilde n}_2}{2 h^2} \left[ - \alpha \, \sin \left( \alpha \, {\tilde \theta} + \beta \, {\tilde \rho}  \right) + r_\Lambda \, \sin \left( {\tilde \theta} \right) \right] = 0 \;, \nonumber\\ 
&& {\tilde \rho}'' + \left( 3 - \frac{{\tilde \theta}^{'2} + {\tilde \rho}^{'2}}{2 \, {\tilde n}_2} \right) {\tilde \rho}'  - \frac{3 {\tilde n}_2}{2 h^2} \, \beta \, \sin \left( \alpha \, {\tilde \theta} + \beta \, {\tilde \rho}  \right) = 0 \;, \nonumber\\ 
&& h' = - h \, \frac{{\tilde \theta}^{'2} + {\tilde \rho}^{'2}}{2 \, {\tilde n}_2} \;,
\end{eqnarray}
where prime denotes derivative w.r.t. ${\hat N}$, and where $h$ is the dimensionless rescaled Hubble rate, $h \equiv H \left( \frac{\sqrt{2}}{\sqrt{3} M_p} \, \frac{\Lambda^2}{\sqrt{1+r_\Lambda}} \right)^{-1}$. 
For definiteness, we evaluate the CMB phenomenology at $N_{\rm CMB} = 60$ e-folds before the end of inflation. Specifically, we evaluate $n_s -1 \simeq - 6 \epsilon + 2 \eta$ and $r \simeq 16 \epsilon$, where, in the two-field context, 
\begin{equation}
\epsilon = \frac{M_p^2}{2} \, \frac{\left\vert \vec{\nabla} V \right\vert^2}{V^2} \;\;\;\; {\rm and} \;\;\;\;
\eta = M_p \, \frac{\dot{\phi}_i \, \dot{\phi}_j}{\dot{\phi}_1^2 + \dot{\phi}_2^2} \, \frac{1}{V} \, \frac{\partial^2 V}{\partial \phi_i \, \partial \phi_j} \;\;\;\; {\rm at} \;\;\;\; \left\{ \phi_1 ,\, \phi_2 \right\} \equiv \left\{ {\hat \theta} ,\, {\hat \rho} \right\} \Big\vert_{N=N_{\rm CMB}} \;.
\label{eps-eta}
\end{equation}

Based on the estimate (\ref{phiN}) - (\ref{phiend1}), we choose initial conditions leading to $70$ e-folds of inflation, so that the fields have settled to the inflationary trajectory (with no ``mistakes'' induced by the slow roll initial approximations) at $N_{\rm CMB}$. To monitor the end of inflation, we evaluate the equation of state $w \equiv p / \rho$ (where $p$ and $\rho$ are, respectively, the pressure and energy density of the two scalars), that using the Einstein equations, is
\begin{equation}
w = - \frac{1}{3} \left[ 1 + \frac{2 \ddot{a}}{H^2 a} \right] = 
- 1 + \frac{{\tilde \theta}^{'2} + {\tilde \rho}^{'2}}{3 \, {\tilde n}_2} \;.
\end{equation}
Inflation occurs for $w < - 1/3$, namely for ${\tilde \theta}^{'2} + {\tilde \rho}^{'2} < 2 \, {\tilde n}_2$.

\begin{figure}
\centering
\includegraphics[scale=0.6]{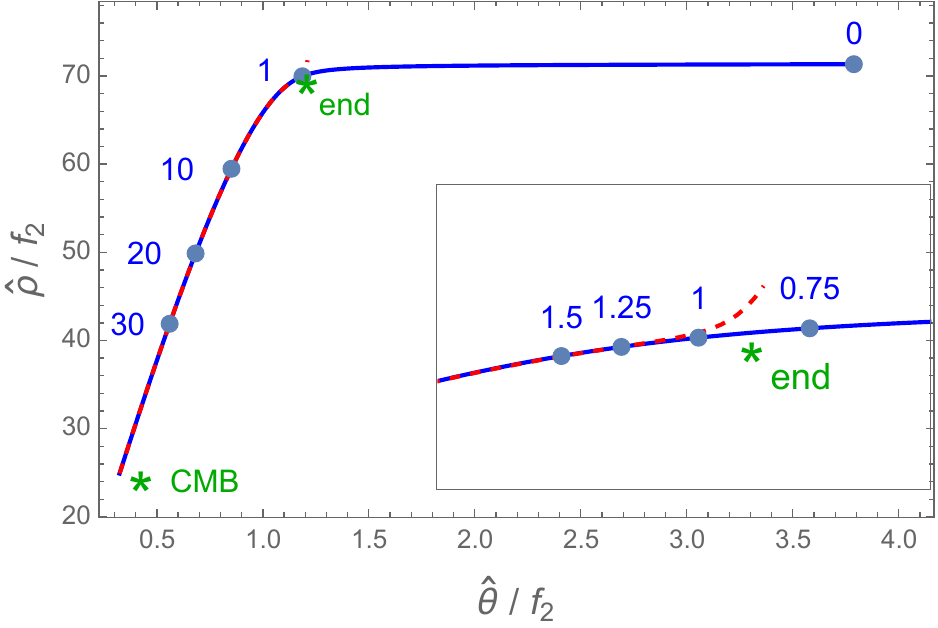}
\caption{Zoom in of the inflationry trajectory shown in Figure~\ref{fig:plotV}. The numerical evolution, shown with a blue curve, is compared with the analytic trajectory, shown with a dashed red curve. Points along the numerical curve indicate the number of e-folds to the end of inflation. The green stars mark our analytic estimate for the position of the fields at the end of inflation and at the CMB production, $60$ e-folds before the end of inflation. The inset shows a further zoom in of the trajectory at $N=1$ e-folds before the end of inflation.
}
\label{fig:traj-ex1}
\end{figure}

The blue curve in Figure~\ref{fig:plotV} shows the final $60$ e-folds of a metastable inflationary trajectory starting from a neighbourhood of the saddle point ${\cal S}_B$. The potential parameters are $n_1 = 160 / M_p^2$, $n_2 = 700 / M_p^2,\, \gamma = 0.0075$, and $r_\Lambda = 0.4$. Figure~\ref{fig:traj-ex1} shows a zoomed-in view of Figure~\ref{fig:plotV} centered in the trajectory. The numerical trajectory, shown in blue, is compared with the analytic trajectory, determined from eq.~(\ref{traj}), and shown with a dashed red curve. The two green points marked with `end' and `CMB' are the analytic estimates for the positions of the two fields at, respectively, the end of the metastable inflationary trajectory (eqs.~(\ref{sol-th-rh})), and $60$ e-folds before the end (the expressions in eq.~(\ref{sol-th-rh}), suppressed by the exponential factor as in eq.~(\ref{phiN}).). The inset shows a further zoom in on the trajectory at $N=1$ e-folds before the end of inflation. 

As we discussed, the inflationary trajectory occurs in the direction of the light eigenmass, in the valley orthogonal to the heavy direction. The two eigenmasses are highly hierarchical at the start of the trajectory (next to the saddle point), and the hierarchy then decreases along the trajectory, until the two masses become comparable, and the trajectory becomes unstable. We plotted the analytic trajectory only up to the point in which the two square eigenmasses become equal (in absolute value). We see that the analytic curve tracks extremely well the numerical one nearly up to this point, which occur at about $1$ e-fold in field space. Our analytic estimate for the end of inflation is very close to the analytic trajectory~\footnote{The small discrepancy is due to the fact that the analytic trajectory is determined from the exact eq.~(\ref{traj}), while the endpoint of inflation is determined by the approximate system~(\ref{end-1}).}, and to the exact position of the fields at about $N=1$ e-fold before the end of inflation. The fields experience a considerable excursion during this final e-fold, however, this discrepancy does not reflect in the CMB estimates, which are essentially off-set only by $1$ e-fold along the $N > 1$ part of the inflationary trajectory. This translates into a small discrepancy in the determination of the field values at CMB scales, and in the evaluation of the slow roll parameters needed for the CMB predictions. Specifically, the numerical evolution leads to 
$n_s = 0.9648$ and to $r = 2.1 \times 10^{-3}$ through eqs.~(\ref{eps-eta}). The analytic result of eqs.~(\ref{ns-r-analytic}) provides a relative discrepancy of about $0.06\%$ for the spectral tilt and of about $3\%$ for the tensor-to-scalar ratio.

\begin{figure}
\centering
\includegraphics[scale=0.45]{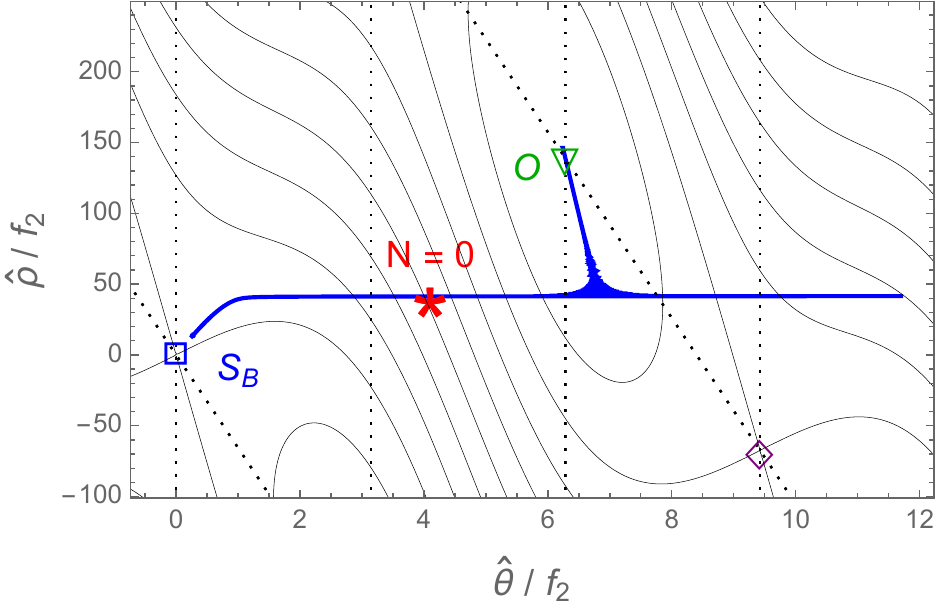}
\includegraphics[scale=0.45]{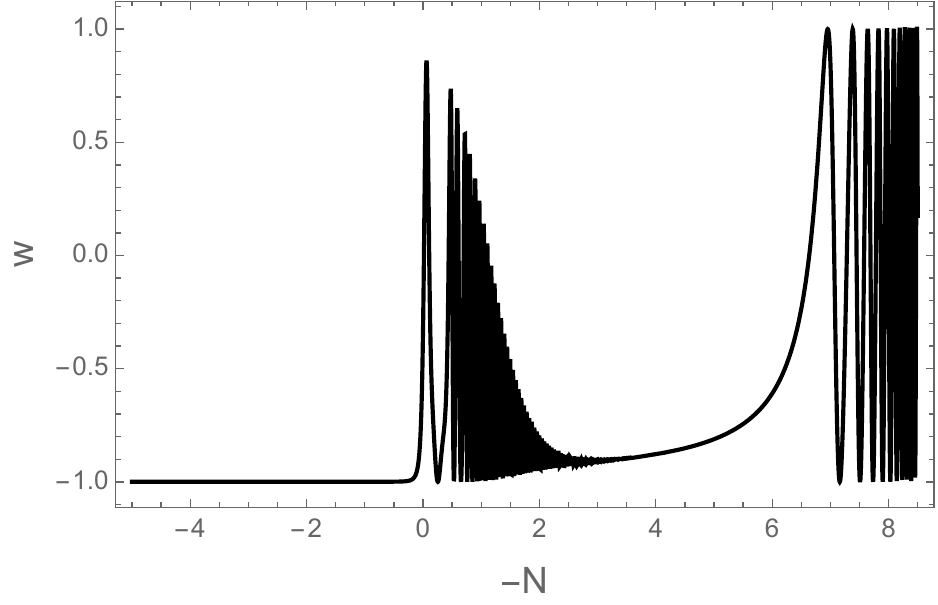}
\caption{Field trajectory (left panel) and equation of state as a function of the number of e-folds (right panel) for a different choice of parameters. The end of the metastable inflationary trajectory (occurring at $N=0$) is followed by oscillations about a new inflationary valley connected to the minimum ${\cal O}$, by a second inflationary stage of duration $\Delta N \simeq 4$ about this valley, and, finally, by oscillations about the minimum. 
}
\label{fig:traj-w--ex2}
\end{figure}

Figure \ref{fig:traj-w--ex2} shows the evolutions of the two fields (left panel) and the equation of state as a function of the number of e-folds (right panel) for a different example, characterized by $n_1 = \frac{70}{M_p^2} ,\, n_2 = \frac{500}{M_p^2} ,\, \gamma = 0.01 ,\, r_\Lambda = 0.25$. Differently from the previous example, the figure shows the evolution also after the end of the metastable inflationary trajectory, that occurs at $N=0$ (marked with a red star in the left panel). Right after this moment, the trajectory reaches the position with greatest value of ${\hat \theta}$ shown in the figure, and then performs oscillations about a next inflationary valley connected to the minimum ${\cal O}$ shown in the figure. When these oscillations are dumped by the expansion, the evolution proceeds along the new inflationary valley, providing an extra amount of inflation of about $\Delta N \simeq 4$. The fields then reach the minimum ${\cal O}$ and start oscillating about it. The right panel shows the corresponding evolution of the equation of state, that is close to $-1$ during the two stages of inflation, and that oscillates while the trajectory is also oscillating. The first oscillation occurs about a negative value, as the minimum value of the potential energy is greater than zero during those oscillations; the second oscillation  occurs about $0$, as it is typical for the oscillations of massive fields. 

This behaviour is markedly different from that of the previous example, in which the evolution after the end of the metastable trajectory (not shown in the two previous figures) was characterized by oscillations about the minimum, without a second inflationary stage. Also in this case, the analytic result (\ref{ns-r-analytic}) for the spectral tilt and the tensor-to-ratio is extremely accurate. Specifically, the value for these two quantities at $N = 60$ e-folds before the end of the first inflationary stage obtained numerically (respectively, analytically) is $n_s = 0.962$ (respectively, $n_s = 0.963$) and $r = 9.6 \times 10^{-4}$ (respectively, $r = 9.4 \times 10^{-4}$). However, the presence of the second inflationary stage shifts the position at which these quantities should be evaluated. For example, evaluating them numerically at $N = 60 - 4 = 56$ e-folds before the end of the metastable trajectory results in $r = 1.1 \times 10^{-3}$ (while $n_s$ does not change at the third significant digit). 

\section{Summary results and CMB phenomenology}
\label{sec:summary}

The model (\ref{knp}) of aligned axion inflation is characterized by $6$ parameters: $2$ potential scales and $4$ axion scales. The scalar spectral tilt and tensor-to-scalar ratio are independent of the overall normalization $\Lambda^4$ of the potential. Furthermore, the redundancy of the model under the internal rotation~(\ref{rotation}) allows to express the results in terms of three combinations of the axions scales, see eq.~(\ref{NC}). We are thus left with the four dimensional parameter space $\left\{ r_\Lambda ,\, n_1 ,\, n_2 ,\, {\cal C} \right\}$. To obtain a simple visualization of the phenomenology of the model, we impose two further constraints: firstly, we fix the degree of alignment by considering a specific value of the dimensionless combination $\gamma$ introduced in eq.~(\ref{gamma}); secondly, we fix the value of the spectral tilt to the best value from CMB observations, $n_s = 0.965$~\cite{Planck:2018jri}. We are thus left with two parameters that we express through the dimensionless combinations
\begin{equation}
r_1 \equiv \frac{n_1}{n_2} \;\;,\;\; {\tilde n}_2 \equiv M_p^2 \, n_2 \;.
\end{equation}

Subject to the constrains on $\gamma$ and $n_s$, the two remaining parameters can be expressed in terms of these ones as
\begin{equation}
{\cal C} = \frac{\gamma \left( 1 + r_1 \right) {\tilde n}_2}{2 M_p^2} \;\;,\;\;
r_\Lambda = \frac{r_1}{1 - q \left( 1 + r_1 \right)^2 {\tilde n}_2} \;\;, 
\label{C-rL-summary}
\end{equation}
where $q$ is the fixed combination
\begin{equation}
q \equiv \frac{\gamma^2}{4 \left( 1 - n_s \right)} \simeq 7 \, \gamma^2 \;.
\end{equation}
A typical degree of alignment of $\gamma = {\rm O } \left( 10^{-2} \right)$ leads to $q = {\rm O } \left( 10^{-3} \right)$. The existence of the metastable trajectory is subject to the conditions~(\ref{cond-metastble}) that, in this parametrization, is satisfied for
\begin{equation}
{\tilde n}_2 \leq \frac{1-\sqrt{r_1}}{q \left( 1 + r_1 \right)^2} \equiv {\tilde n}_{2,{\rm max}} \left( r_1 \right) \;.
\label{cond-metastble-2}
\end{equation}

\begin{figure}
\centering
\includegraphics[scale=0.4]{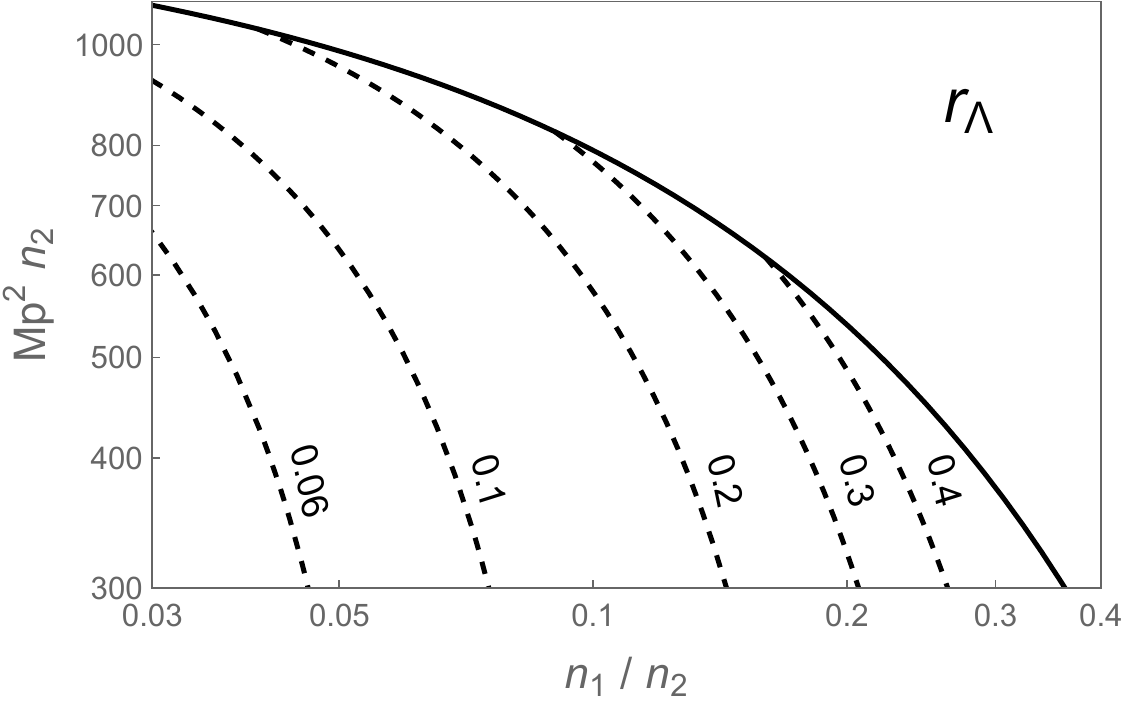}
\includegraphics[scale=0.4]{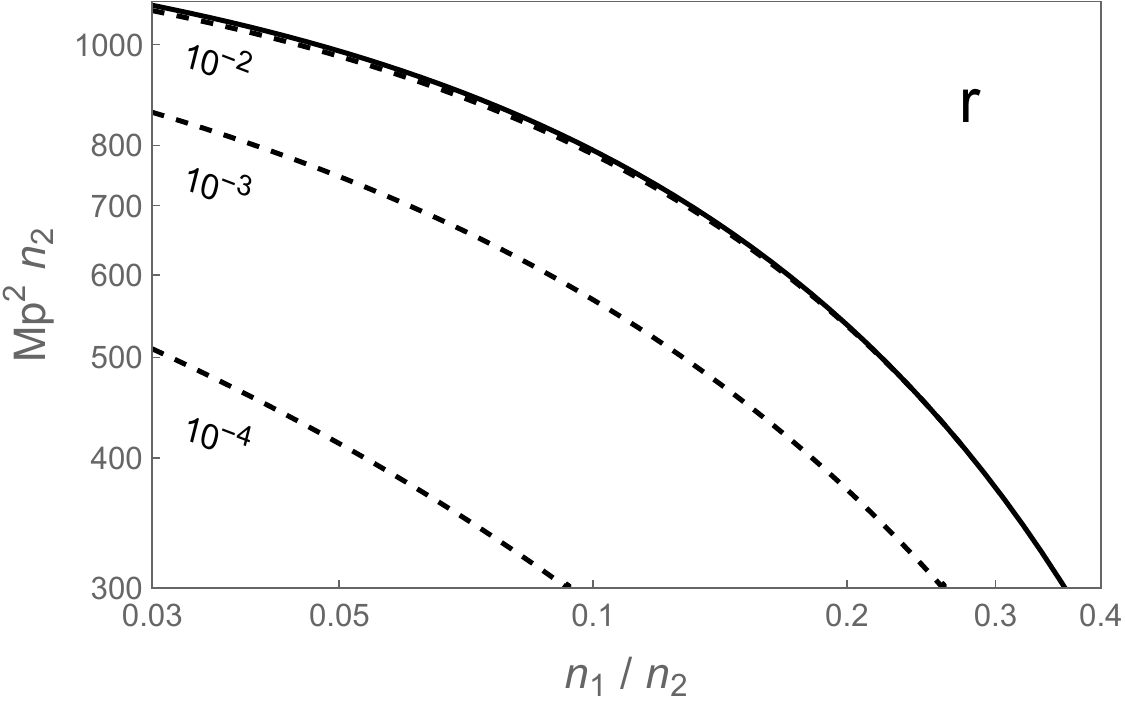}
\caption{Left panel: ratio $r_\Lambda$ between the two potential scales (see eq.~(\ref{Lambda-resca})); right panel: 
tensor to scalar ratio $r$. These quantities are shown as a  function of two model parameters in the region below the solid curve, determined by the condition~(\ref{cond-metastble-2}) of existence of the metastable inflationary trajectory. The results shown are subject to the constraints of an alignment parameter $\gamma = 10^{-2}$ and of a scalar spectral tilt $n_s = 0.965$. The tensor-to-scalar ratio is quantified disregarding the possible existence of a second inflationary trajectory, see the text for details.}
\label{fig:summaryCMB}
\end{figure}

In the left panel of Figure~\ref{fig:summaryCMB}, we show the ratio $r_\Lambda$ of the two potential scales, as determined by the second of~(\ref{C-rL-summary}). In the right panel we show instead the tensor-to-scalar-ratio, as determined by the second of~(\ref{ns-r-analytic}). In the full region shown in the figure, the tensor-to-scalar ratio is below the current bound $r \la 0.03$~\cite{Tristram:2021tvh,Galloni:2022mok}, reaching its maximum value $r \simeq 0.02$ in the top left corner of the figure. Inserting the expressions~(\ref{C-rL-summary}) into~(\ref{ns-r-analytic}), we find that $r = {\rm O} \left( {\tilde n}_2^2 \right)$ at small ${\tilde n}_2$. We also saw numerically that, at fixed $r_1$, the tensor-to-scalar ratio is a growing function of ${\tilde n}_2$ in the full domain. At the upper boundary~(\ref{cond-metastble-2}), we obtain
\begin{equation}
r \left( r_1 ,\, {\tilde n}_{2,{\rm max}} \left( r_1 \right) \right) = \frac{\pi^2 \, \left\vert n_s - 1 \right\vert}{2} \, \left( 1 - \sqrt{r_1} \right) \, {\rm e}^{-\left\vert n_s - 1 \right\vert N} \simeq 0.021 \left( 1 - \sqrt{r_1} \right) \,,
\label{r-upper-boundary}
\end{equation}
where $n_s = 0.965$ and $N = 60$ have been assumed in the last step. We thus see that the maximum value of the tensor-to-scalar ratio is obtained in the top left corner of the allowed domain, in agreement with what observed in Figure~\ref{fig:summaryCMB}. Part of this parameter space will be probed in the near future by CMB polarization measurements such as CMB-S4~\cite{CMB-S4:2020lpa} and LiteBIRD~\cite{LiteBIRD:2023iei}, that are expected to reach an ${\rm O } \left( 10^{-3} \right)$ sensitivity on the tensor-to-scalar ratio.

From these results we can also easily estimate the scale of inflation $\Lambda$. This is fixed by requiring that the amplitude of the scalar perturbations, evaluated in slow roll, matches the observed one~\cite{Planck:2018vyg}
\begin{equation}
A_s \simeq \left( \frac{H}{\dot{\phi}} \right)^2 \left( \frac{H}{2 \pi} \right)^2 \simeq \frac{V}{24 \pi^2 \, M_p^4  \epsilon} \simeq 2.1 \times 10^{-9} \;.
\end{equation}
To estimate the scale of the potential $\Lambda$, we ``trade'' $\epsilon$ for $r/16$ and we evaluate $V$ at the saddle point, obtaining
\begin{equation}
\Lambda \simeq 2.7 \times 10^{16} \, {\rm GeV} \, \times r^{1/4} \left( 1 + r_\Lambda \right)^{1/4} \simeq 5 \times 10^{15} \, {\rm GeV} \left( \frac{r}{10^{-3}} \right)^{1/4} \;, 
\label{scale-V}
\end{equation}
where $\left( 1 + r_\Lambda \right)^{1/4}$ has been approximated to $1$ in the final estimate. 

We conclude by cautioning the reader that the tensor-to-scalar ratio result shown in Figure~\ref{fig:summaryCMB} assumes $60$ e-folds of inflation in the metastable trajectory. We showed in the previous section that, depending on the model parameters, the fields could reach a second inflationary trajectory connected to a minimum. When this happens the number of e-folds of this second trajectory denoted by $N_2$ should be subtracted from the total number of e-folds in the evaluation of the second of ~(\ref{C-rL-summary}), as only $N_1 = N_{\rm CMB} - N_2$ e-folds of inflation occur in the metastable trajectory from the horizon exit of the CMB modes to its end.

\section{The trans-Planckian problem}
\label{sec:transPlanck}

After having obtained the CMB phenomenology associated with the metastable trajectories streaming off the saddle points, we can now move to a more theoretical consideration on the viable range of parameters. Substituting $n_s=0.965$ and $N=60$ in the second of (\ref{phiN}), we obtain the estimate
\begin{equation}
\frac{\Delta \phi}{M_p} \simeq \sqrt{\frac{r}{0.02}} \;.
\end{equation}
for the displacement of the inflaton along the metastable inflationary trajectory. Figure~\ref{fig:summaryCMB} and eq.~(\ref{r-upper-boundary}) indicate that this excursion is typically sub-Planckian.

However, as discussed in the introduction, a number of works in the literature have questioned the stability of the model against quantum gravity corrections. As done in Refs.~\cite{Rudelius:2015xta,Montero:2015ofa}, we evaluate the eigenvalues of the axion kinetic matrix, for the range of parameters leading to a correct phenomenology associated to the metastable inflationary trajectory. To this goal, we introduce the so called {\it lattice basis}
\begin{equation}
\chi_1 \equiv f_1^{-1} \, \theta + g_1^{-1} \rho \;\;\;,\;\;\; \chi_2 \equiv f_2^{-1} \, \theta + g_2^{-1} \rho \;, 
\end{equation}
in terms of which the lagrangian of the model acquires the form 
\begin{equation}
{\cal L} = - \frac{1}{2} \sum_{i=1,2} {\cal K}_{ij} \, \partial \chi_i \, \partial \chi_j - \Lambda_1^4 \left( 1 - \cos \, \chi_1 \right) - \Lambda_2^4 \left( 1 - \cos \, \chi_2 \right) \;,
\end{equation}
with a diagonal potential and a non-diagonal kinetic term. The kinetic matrix has eigenvalues
\begin{equation}
\kappa_\pm^2 \equiv \frac{n_1+n_2 \pm \sqrt{\left( n_1 + n_2 \right)^2 - 4 \, {\cal C}^2}}{2 \, {\cal C}^2} \;,
\end{equation}
which are obviously invariant under the rotation~(\ref{rotation}). In this basis, the alignment mechanism of~\cite{Kim:2004rp} is encoded in the fact that $\kappa_+$ is enhanced in the limit of small alignment,
\begin{equation}
\lim_{\left\vert \gamma \right\vert \ll 1} \kappa_+^2 = \frac{n_1+n_2}{{\cal C}^2} \;.
\end{equation}

As we now show, our result for the scalar spectral tilt alone implies that this eigenvalue is trans-Planckian. From the first of~(\ref{ns-r-analytic}), we indeed obtain
\begin{equation}
\frac{\kappa_+^2}{M_p^2} \simeq 
\frac{1}{1 - n_s} \times \frac{r_\Lambda \left( n_2 + n_1 \right)}{r_\Lambda \, n_2 - n_1} \;\;,
\end{equation}
as the second factor is generally greater than $1$, and it approaches $1$ for $n_1 \ll r_\Lambda n_2$, we then see that 
\begin{equation}
\frac{\kappa_+}{M_p} \ga \frac{1}{\left\vert n_s - 1 \right\vert^{1/2}} \simeq 5.3 \;,
\label{k+res}
\end{equation}
where the CMB result $n_s \simeq 0.965$~\cite{Planck:2018jri} has been used in the last step. We therefore see that, even if we account for the metastable solutions, the minimal implementation of aligned axion inflation model~\cite{Kim:2004rp} is not compatible with the WGC, and one need to invoke additional ingredients to solve this problem such as the existence of additional instantons, as we have discussed in the introduction.

\section{Conclusions}
\label{sec:concl}

Most of the studies of aligned axion inflation~\cite{Kim:2004rp} have focused on the consistency of the model with quantum gravity. In most of the analyses, the heavy direction is integrated out, and inflation is assumed to proceed along a trajectory that terminates close to a minimum of the two-field potential, along which the trajectory resembles that of the single field original model of natural inflation~\cite{Freese:1990rb}. It was shown in~\cite{Peloso:2015dsa} that these solutions provide a tensor-to-scalar ratio larger than that of~\cite{Freese:1990rb}. Therefore, analogously to this model, they are ruled out by CMB data~\cite{Planck:2018jri}. Compatibility with observations can be obtained in presence of small modulation from higher-order instantons~\cite{Kappl:2015esy,Choi:2015aem}. These instantons might also help reconciling the model with the WGC~\cite{Bachlechner:2015qja,Hebecker:2015rya,Brown:2015iha,Brown:2015lia}. Moreover, it was noted in~\cite{Peloso:2015dsa} that the original implementation of~\cite{Kim:2004rp} possesses a second class of solutions, that stream off saddle points of the two-field potential, and terminate away from the minima of the potential, due to an instability in the orthogonal direction. While these solutions were studied only numerically in~\cite{Peloso:2015dsa}, the present work constitutes a complete analytic study of these trajectories and of their associated phenomenology. 

In addition to~\cite{Peloso:2015dsa}, we: (1) obtained an analytic condition, eq.~(\ref{traj}), that determines the metastable trajectories; this condition acquires a very compact form in the limit of strong alignment, see eq.~(\ref{trajectory}); (2) showed that also the condition for the stability of the trajectory acquires a very simple form in the limit of strong alignment, see eq.~(\ref{stability}); the combination of these two conditions allows to obtain an analytic expression for the end-point of inflation, see eqs.~(\ref{sol-th-rh}), which, as we showed in Section~\ref{sec:numerical-CMB}, agrees with the exact numerical integration with a mistake of less than about one e-fold, see Figure~\ref{fig:traj-ex1} for one example; (3) used this result to obtain extremely simple and accurate analytic expressions for the scalar spectral tilt and the tensor-to-scalar ratio, see eqs.~(\ref{ns-r-analytic}). 

The result~(\ref{ns-r-analytic}) allows for an exhaustive analytic study of the CMB phenomenology of these trajectories as a function of three independent model parameters. We formulated our study in a basis-independent manner, exploiting the invariance of the model under internal rotations between the two axions. This allows to formulate the model in terms of three axion scales, that we parametrized via the three invariant combinations~(\ref{NC}), and of the two potential scales, that we parametrize through their sum and ratio in eq.~(\ref{Lambda-resca}). As standard in inflation, the sum can be expressed through the other parameters from the amplitude of scalar perturbations, see eq.~(\ref{scale-V}). We further reduce the parameter space by fixing the scalar spectral tilt to the CMB best value $n_s = 0.965$~\cite{Planck:2018jri}. We end up with a three-dimensional parameter space. We can further reduce the dimension by fixing the degree of alignment, defined via eq.~(\ref{gamma}) in a basis-independent way. This allows to present the phenomenology of the model with two-dimensional contour plots, see Figure~\ref{fig:summaryCMB}.

From the analytic result~(\ref{ns-r-analytic}) of the scalar spectral tilt, we could also precisely quantify the eigenvalues of the kinetic matrix of the model in the so called lattice basis, in which the potential is diagonal. Eq.~(\ref{k+res}) immediately shows that one eigenvalue is trans-Plankian, therefore also the metastable trajectories suffer from the problem pointed out in~\cite{Rudelius:2015xta,Montero:2015ofa}, for which a solution along the lines of~\cite{Bachlechner:2015qja,Hebecker:2015rya,Brown:2015iha,Brown:2015lia} might possibly exist. 

Finally, it is instructive, and in some cases phenomenologically relevant, to study the evolution of the two fields after the instability that terminates the inflationary trajectory. Ref.~\cite{Peloso:2015dsa} presented one example in which the two fields reached a minimum of the potential right after this instability, see their Figure~3. In our numerical investigations, we encountered a new class of solutions, for which, after the instability, the fields oscillates about a new inflationary valley, connected to a minimum of the potential, about a point away from the minimum. These oscillations are rapidly damped by the expansion, and then a second inflationary evolution takes place along this valley, see Figure~\ref{fig:traj-w--ex2}. Whether this second stage of inflation takes place or not, and its duration, depend on whether the immediate evolution of the two fields after the instability is close or far from a minimum (compare Figure~3 of~\cite{Peloso:2015dsa} with our Figure~\ref{fig:traj-w--ex2}). Answering this question goes beyond the scope of the present work, that focuses on the analytic study of the metastable part of the trajectory, but we believe it deserves a separate investigation, since the additional amount of e-folds of the second inflationary stage (when present) need to be included when evaluating the CMB phenomenology of the model. 

Multiple-field models of inflation, and particularly multiple axion models, have been extensively studied in the literature, and we hope that some of the the analytic tools developed in this work, and some of the observations emerged from our study (for example the presence of an inflationary trajectory next to critical points and of multiple stages of inflation), can be of use also beyond the specific model studied in this work. This includes for instance: Ref.~\cite{McAllister:2012am}, that considered inflationary trajectories near inflection points in multi-field models that can be described by a random matrix model; Ref.~\cite{Battefeld:2013xwa}, that studied inflation near saddle points, provided an effective description for the end of these trajectories; Ref.~\cite{Carta:2020oci}, that formulated a realization of hybrid inflation with a trajectory that also starts from a saddle point in a two-axion model; Refs.~\cite{Chatzistavrakidis:2012bb,Choi:2014rja} that considered the mechanism of aligned axion inflation with more than two axions. We believe that our results might be useful also to extend the studies of these models, and we hope to come back to this in some future work.

\section*{Acknowledgements}
We are grateful to Luca Martucci for useful discussions. We acknowledge support from Istituto Nazionale di Fisica Nucleare (INFN)
through the Theoretical Astroparticle Physics (TAsP) project, and from the MIUR Progetti di Ricerca
di Rilevante Interesse Nazionale (PRIN) Bando 2022 - grant 20228RMX4A.

\appendix

\section{Eigenmasses, eigenvectors, and the inflationary trajectory}
\label{app:eigensystem}

The squared eigenmasses of the potenatial in~(\ref{potential}) are the eigenvalues of the Hessian, namely of the $2 \times 2$ matrix
\begin{equation}
V''_{ij} \equiv \frac{\partial^2 V}{\partial \phi_i \, \partial \phi_j} \;\;\;,\;\;\; \phi_i \equiv \left\{ {\hat \theta} ,\, {\hat \rho} \right\}_i \;.
\end{equation}
A direct computation gives the two eigenvalues
\begin{eqnarray}
m_\pm^2 &=& \frac{\Lambda^4}{2 \left( 1 + r_\Lambda \right)} 
\Bigg[ -n_1 \, \cos {\cal A}_1 + r_\Lambda \, n_2 \, \cos {\cal A}_2   \nonumber\\
&& \quad\quad  \quad\quad \pm \sqrt{\left( -n_1 \, \cos {\cal A}_1 + r_\Lambda \, n_2 \, \cos {\cal A}_2  \right)^2 - 4 r_\Lambda \, {\cal C}^2 \cos {\cal A}_1 \cos {\cal A}_2} \Bigg] \nonumber\\
&\equiv& \frac{\Lambda^4}{2 \left( 1 + r_\Lambda \right)} \, {\tilde m}_\pm^2 \;.
\label{eigen-mass}
\end{eqnarray}
The corresponding unit-norm eigenvectors are
\begin{equation}
\vec{v}_\pm = {\cal N}_\pm \left\{ {\tilde m}_\pm^2 + 2 \left( g_1^{-2} \cos {\cal A}_1 - g_2^{-2} r_\Lambda \cos {\cal A}_2 \right) ,\, 2 \left( -f_1^{-1} g_1^{-1} \cos {\cal A}_1 + f_2^{-1} g_2^{-1} r_\Lambda \cos {\cal A}_2 \right) \right\}  \;,
\label{eigen-vector}
\end{equation}
where the normalization factors are
\begin{equation}
{\cal N}_\pm \equiv \frac{1}{\sqrt{2}} \left[ {\tilde m}_\pm^2 + n_1 \cos {\cal A}_1 - r_\Lambda n_2 {\cal A}_2 \right]^{-1/2} \left[ {\tilde m}_\pm^2 + 2 \left( g_1^{-2} \cos {\cal A}_1 - g_2^{-2} r_\Lambda \cos {\cal A}_2 \right)  \right]^{-1/2} \;. 
\end{equation}

The expressions (\ref{eigen-mass}) and (\ref{eigen-vector}) are valid for a generic point in field space, and they are exact in the model parameters. Evaluating the two expressions (\ref{eigen-mass}) at the critical points gives the squared eigenmasses reported in the last column of Table~\ref{tab:critical}. In Section~\ref{sec:trajectory}, we study two inflationary trajectories. For both cases $n_1 \, \cos {\cal A}_1 + r_\Lambda \, n_2 \, \cos {\cal A}_2 > 0$ at the corresponding saddle point, so that the positive squared eigenmass is $m_+^2$ for both cases. 

The condition (\ref{def-traj}) for the trjectory then evaluates to
\begin{eqnarray}
0 = \vec{v}_+ \cdot \vec{\nabla} V &=& \frac{\Lambda^4}{1 + r_\Lambda} {\cal N}_+
\Bigg\{ \left[ {\tilde m}_+^2 - 2 \left( g_1^{-2} \cos {\cal A}_1 - g_2^{-2} r_\Lambda \cos {\cal A}_2 \right) \right] \left( -f_1^{-1} \sin {\cal A}_1 + f_2^{-1} \sin {\cal A}_2 \right) \nonumber\\
&& \quad\quad  \quad\quad  \quad + 2 \left( f_1^{-1} g_1^{-1} \cos {\cal A}_1 - f_2^{-1} g_2^{-1} r_\Lambda \cos {\cal A}_2 \right) \left( g_1^{-1} \sin {\cal A}_1 - g_2^{-1} \sin {\cal A}_2 \right) \Bigg\} \;. \nonumber\\ 
\label{traj}
\end{eqnarray}

In the limit of strong alignment, the two expressions (\ref{eigen-mass}) and (\ref{traj}) can be simplified by setting ${\cal C} = 0$ in the coefficients outside ${\cal A}_{1,2}$. Namely, we set 
\begin{equation}
f_2^{-1} = \sqrt{\frac{n_2}{n_1}} \, f_1^{-1} \;\;,\;\; 
g_1^{-1}  = \sqrt{n_1 - f_1^{-2}} \;\;,\;\; 
g_2^{-1} = \sqrt{\frac{n_2 \left( n_1-f_1^{-2} \right)}{n_1}} \;. 
\end{equation}

Doing so, eq. (\ref{traj}) simplifies to 
\begin{eqnarray}
\vec{v}_+ \cdot \vec{\nabla} V &=& \frac{\Lambda^4}{1 + r_\Lambda} {\cal N}_+
{\tilde m}_+^2 \frac{f_1^{-1}}{\sqrt{n_1}} \left( -\sqrt{n_1} \, \sin {\cal A}_1 + r_\Lambda \, \sqrt{n_2} \sin {\cal A}_2 \right) = 0 \;,
\end{eqnarray}
from which the condition (\ref{trajectory}) of the main text results. Performing the same simplification in eq.~(\ref{eigen-mass}) gives
\begin{equation}
m_+^2 \simeq 2 \left( -n_1 \, \cos {\cal A}_1 + r_\Lambda \, n_2 \, \cos {\cal A}_2 \right) \;. 
\end{equation}
from which the condition (\ref{stability}) of the main text results.

\bibliographystyle{jhep}
\bibliography{bibliography}

\providecommand{\href}[2]{#2}\begingroup\raggedright\begin{thebibliography}{10}

\bibitem{Guth:1980zm}
A.H.~Guth, \emph{{The Inflationary Universe: A Possible Solution to the Horizon
  and Flatness Problems}},
  \href{https://doi.org/10.1103/PhysRevD.23.347}{\emph{Phys. Rev. D} {\bfseries
  23} (1981) 347}.

\bibitem{Linde:1982uu}
A.D.~Linde, \emph{{Scalar Field Fluctuations in Expanding Universe and the New
  Inflationary Universe Scenario}},
  \href{https://doi.org/10.1016/0370-2693(82)90293-3}{\emph{Phys. Lett. B}
  {\bfseries 116} (1982) 335}.

\bibitem{Albrecht:1982wi}
A.~Albrecht and P.J.~Steinhardt, \emph{{Cosmology for Grand Unified Theories
  with Radiatively Induced Symmetry Breaking}},
  \href{https://doi.org/10.1103/PhysRevLett.48.1220}{\emph{Phys. Rev. Lett.}
  {\bfseries 48} (1982) 1220}.

\bibitem{Planck:2018jri}
{\scshape Planck} collaboration, \emph{{Planck 2018 results. X. Constraints on
  inflation}}, \href{https://doi.org/10.1051/0004-6361/201833887}{\emph{Astron.
  Astrophys.} {\bfseries 641} (2020) A10}
  [\href{https://arxiv.org/abs/1807.06211}{{\ttfamily 1807.06211}}].

\bibitem{BICEP:2021xfz}
{\scshape BICEP, Keck} collaboration, \emph{{Improved Constraints on Primordial
  Gravitational Waves using Planck, WMAP, and BICEP/Keck Observations through
  the 2018 Observing Season}},
  \href{https://doi.org/10.1103/PhysRevLett.127.151301}{\emph{Phys. Rev. Lett.}
  {\bfseries 127} (2021) 151301}
  [\href{https://arxiv.org/abs/2110.00483}{{\ttfamily 2110.00483}}].

\bibitem{Lyth:1996im}
D.H.~Lyth, \emph{{What would we learn by detecting a gravitational wave signal
  in the cosmic microwave background anisotropy?}},
  \href{https://doi.org/10.1103/PhysRevLett.78.1861}{\emph{Phys. Rev. Lett.}
  {\bfseries 78} (1997) 1861}
  [\href{https://arxiv.org/abs/hep-ph/9606387}{{\ttfamily hep-ph/9606387}}].

\bibitem{Freese:1990rb}
K.~Freese, J.A.~Frieman and A.V.~Olinto, \emph{{Natural inflation with pseudo -
  Nambu-Goldstone bosons}},
  \href{https://doi.org/10.1103/PhysRevLett.65.3233}{\emph{Phys. Rev. Lett.}
  {\bfseries 65} (1990) 3233}.

\bibitem{Adams:1992bn}
F.C.~Adams, J.R.~Bond, K.~Freese, J.A.~Frieman and A.V.~Olinto, \emph{{Natural
  inflation: Particle physics models, power law spectra for large scale
  structure, and constraints from COBE}},
  \href{https://doi.org/10.1103/PhysRevD.47.426}{\emph{Phys. Rev. D} {\bfseries
  47} (1993) 426} [\href{https://arxiv.org/abs/hep-ph/9207245}{{\ttfamily
  hep-ph/9207245}}].

\bibitem{Pajer:2013fsa}
E.~Pajer and M.~Peloso, \emph{{A review of Axion Inflation in the era of
  Planck}}, \href{https://doi.org/10.1088/0264-9381/30/21/214002}{\emph{Class.
  Quant. Grav.} {\bfseries 30} (2013) 214002}
  [\href{https://arxiv.org/abs/1305.3557}{{\ttfamily 1305.3557}}].

\bibitem{Ghigna:1992iv}
S.~Ghigna, M.~Lusignoli and M.~Roncadelli, \emph{{Instability of the invisible
  axion}}, \href{https://doi.org/10.1016/0370-2693(92)90019-Z}{\emph{Phys.
  Lett. B} {\bfseries 283} (1992) 278}.

\bibitem{Holman:1992us}
R.~Holman, S.D.H.~Hsu, T.W.~Kephart, E.W.~Kolb, R.~Watkins and L.M.~Widrow,
  \emph{{Solutions to the strong CP problem in a world with gravity}},
  \href{https://doi.org/10.1016/0370-2693(92)90491-L}{\emph{Phys. Lett. B}
  {\bfseries 282} (1992) 132}
  [\href{https://arxiv.org/abs/hep-ph/9203206}{{\ttfamily hep-ph/9203206}}].

\bibitem{Kamionkowski:1992mf}
M.~Kamionkowski and J.~March-Russell, \emph{{Planck scale physics and the
  Peccei-Quinn mechanism}},
  \href{https://doi.org/10.1016/0370-2693(92)90492-M}{\emph{Phys. Lett. B}
  {\bfseries 282} (1992) 137}
  [\href{https://arxiv.org/abs/hep-th/9202003}{{\ttfamily hep-th/9202003}}].

\bibitem{Giddings:1987cg}
S.B.~Giddings and A.~Strominger, \emph{{Axion Induced Topology Change in
  Quantum Gravity and String Theory}},
  \href{https://doi.org/10.1016/0550-3213(88)90446-4}{\emph{Nucl. Phys. B}
  {\bfseries 306} (1988) 890}.

\bibitem{Giddings:1989bq}
S.B.~Giddings and A.~Strominger, \emph{{STRING WORMHOLES}},
  \href{https://doi.org/10.1016/0370-2693(89)91651-1}{\emph{Phys. Lett. B}
  {\bfseries 230} (1989) 46}.

\bibitem{Kallosh:1995hi}
R.~Kallosh, A.D.~Linde, D.A.~Linde and L.~Susskind, \emph{{Gravity and global
  symmetries}}, \href{https://doi.org/10.1103/PhysRevD.52.912}{\emph{Phys. Rev.
  D} {\bfseries 52} (1995) 912}
  [\href{https://arxiv.org/abs/hep-th/9502069}{{\ttfamily hep-th/9502069}}].

\bibitem{Banks:2003sx}
T.~Banks, M.~Dine, P.J.~Fox and E.~Gorbatov, \emph{{On the possibility of large
  axion decay constants}},
  \href{https://doi.org/10.1088/1475-7516/2003/06/001}{\emph{JCAP} {\bfseries
  06} (2003) 001} [\href{https://arxiv.org/abs/hep-th/0303252}{{\ttfamily
  hep-th/0303252}}].

\bibitem{Dimopoulos:2005ac}
S.~Dimopoulos, S.~Kachru, J.~McGreevy and J.G.~Wacker, \emph{{N-flation}},
  \href{https://doi.org/10.1088/1475-7516/2008/08/003}{\emph{JCAP} {\bfseries
  08} (2008) 003} [\href{https://arxiv.org/abs/hep-th/0507205}{{\ttfamily
  hep-th/0507205}}].

\bibitem{Rudelius:2014wla}
T.~Rudelius, \emph{{On the Possibility of Large Axion Moduli Spaces}},
  \href{https://doi.org/10.1088/1475-7516/2015/04/049}{\emph{JCAP} {\bfseries
  04} (2015) 049} [\href{https://arxiv.org/abs/1409.5793}{{\ttfamily
  1409.5793}}].

\bibitem{McAllister:2008hb}
L.~McAllister, E.~Silverstein and A.~Westphal, \emph{{Gravity Waves and Linear
  Inflation from Axion Monodromy}},
  \href{https://doi.org/10.1103/PhysRevD.82.046003}{\emph{Phys. Rev. D}
  {\bfseries 82} (2010) 046003}
  [\href{https://arxiv.org/abs/0808.0706}{{\ttfamily 0808.0706}}].

\bibitem{Anber:2009ua}
M.M.~Anber and L.~Sorbo, \emph{{Naturally inflating on steep potentials through
  electromagnetic dissipation}},
  \href{https://doi.org/10.1103/PhysRevD.81.043534}{\emph{Phys. Rev. D}
  {\bfseries 81} (2010) 043534}
  [\href{https://arxiv.org/abs/0908.4089}{{\ttfamily 0908.4089}}].

\bibitem{Adshead:2012kp}
P.~Adshead and M.~Wyman, \emph{{Chromo-Natural Inflation: Natural inflation on
  a steep potential with classical non-Abelian gauge fields}},
  \href{https://doi.org/10.1103/PhysRevLett.108.261302}{\emph{Phys. Rev. Lett.}
  {\bfseries 108} (2012) 261302}
  [\href{https://arxiv.org/abs/1202.2366}{{\ttfamily 1202.2366}}].

\bibitem{Kim:2004rp}
J.E.~Kim, H.P.~Nilles and M.~Peloso, \emph{{Completing natural inflation}},
  \href{https://doi.org/10.1088/1475-7516/2005/01/005}{\emph{JCAP} {\bfseries
  01} (2005) 005} [\href{https://arxiv.org/abs/hep-ph/0409138}{{\ttfamily
  hep-ph/0409138}}].

\bibitem{Rudelius:2015xta}
T.~Rudelius, \emph{{Constraints on Axion Inflation from the Weak Gravity
  Conjecture}}, \href{https://doi.org/10.1088/1475-7516/2015/9/020}{\emph{JCAP}
  {\bfseries 09} (2015) 020}
  [\href{https://arxiv.org/abs/1503.00795}{{\ttfamily 1503.00795}}].

\bibitem{Montero:2015ofa}
M.~Montero, A.M.~Uranga and I.~Valenzuela, \emph{{Transplanckian axions!?}},
  \href{https://doi.org/10.1007/JHEP08(2015)032}{\emph{JHEP} {\bfseries 08}
  (2015) 032} [\href{https://arxiv.org/abs/1503.03886}{{\ttfamily
  1503.03886}}].

\bibitem{Brown:2015iha}
J.~Brown, W.~Cottrell, G.~Shiu and P.~Soler, \emph{{Fencing in the Swampland:
  Quantum Gravity Constraints on Large Field Inflation}},
  \href{https://doi.org/10.1007/JHEP10(2015)023}{\emph{JHEP} {\bfseries 10}
  (2015) 023} [\href{https://arxiv.org/abs/1503.04783}{{\ttfamily
  1503.04783}}].

\bibitem{Bachlechner:2015qja}
T.C.~Bachlechner, C.~Long and L.~McAllister, \emph{{Planckian Axions and the
  Weak Gravity Conjecture}},
  \href{https://doi.org/10.1007/JHEP01(2016)091}{\emph{JHEP} {\bfseries 01}
  (2016) 091} [\href{https://arxiv.org/abs/1503.07853}{{\ttfamily
  1503.07853}}].

\bibitem{Hebecker:2015rya}
A.~Hebecker, P.~Mangat, F.~Rompineve and L.T.~Witkowski, \emph{{Winding out of
  the Swamp: Evading the Weak Gravity Conjecture with F-term Winding
  Inflation?}},
  \href{https://doi.org/10.1016/j.physletb.2015.07.026}{\emph{Phys. Lett. B}
  {\bfseries 748} (2015) 455}
  [\href{https://arxiv.org/abs/1503.07912}{{\ttfamily 1503.07912}}].

\bibitem{Brown:2015lia}
J.~Brown, W.~Cottrell, G.~Shiu and P.~Soler, \emph{{On Axionic Field Ranges,
  Loopholes and the Weak Gravity Conjecture}},
  \href{https://doi.org/10.1007/JHEP04(2016)017}{\emph{JHEP} {\bfseries 04}
  (2016) 017} [\href{https://arxiv.org/abs/1504.00659}{{\ttfamily
  1504.00659}}].

\bibitem{Heidenreich:2015wga}
B.~Heidenreich, M.~Reece and T.~Rudelius, \emph{{Weak Gravity Strongly
  Constrains Large-Field Axion Inflation}},
  \href{https://doi.org/10.1007/JHEP12(2015)108}{\emph{JHEP} {\bfseries 12}
  (2015) 108} [\href{https://arxiv.org/abs/1506.03447}{{\ttfamily
  1506.03447}}].

\bibitem{Kappl:2015esy}
R.~Kappl, H.P.~Nilles and M.W.~Winkler, \emph{{Modulated Natural Inflation}},
  \href{https://doi.org/10.1016/j.physletb.2015.12.073}{\emph{Phys. Lett. B}
  {\bfseries 753} (2016) 653}
  [\href{https://arxiv.org/abs/1511.05560}{{\ttfamily 1511.05560}}].

\bibitem{Choi:2015aem}
K.~Choi and H.~Kim, \emph{{Aligned natural inflation with modulations}},
  \href{https://doi.org/10.1016/j.physletb.2016.05.097}{\emph{Phys. Lett. B}
  {\bfseries 759} (2016) 520}
  [\href{https://arxiv.org/abs/1511.07201}{{\ttfamily 1511.07201}}].

\bibitem{Arkani-Hamed:2006emk}
N.~Arkani-Hamed, L.~Motl, A.~Nicolis and C.~Vafa, \emph{{The String landscape,
  black holes and gravity as the weakest force}},
  \href{https://doi.org/10.1088/1126-6708/2007/06/060}{\emph{JHEP} {\bfseries
  06} (2007) 060} [\href{https://arxiv.org/abs/hep-th/0601001}{{\ttfamily
  hep-th/0601001}}].

\bibitem{Cheung:2014vva}
C.~Cheung and G.N.~Remmen, \emph{{Naturalness and the Weak Gravity
  Conjecture}},
  \href{https://doi.org/10.1103/PhysRevLett.113.051601}{\emph{Phys. Rev. Lett.}
  {\bfseries 113} (2014) 051601}
  [\href{https://arxiv.org/abs/1402.2287}{{\ttfamily 1402.2287}}].

\bibitem{Long:2014dta}
C.~Long, L.~McAllister and P.~McGuirk, \emph{{Aligned Natural Inflation in
  String Theory}},
  \href{https://doi.org/10.1103/PhysRevD.90.023501}{\emph{Phys. Rev. D}
  {\bfseries 90} (2014) 023501}
  [\href{https://arxiv.org/abs/1404.7852}{{\ttfamily 1404.7852}}].

\bibitem{Gao:2014uha}
X.~Gao, T.~Li and P.~Shukla, \emph{{Combining Universal and Odd RR Axions for
  Aligned Natural Inflation}},
  \href{https://doi.org/10.1088/1475-7516/2014/10/048}{\emph{JCAP} {\bfseries
  10} (2014) 048} [\href{https://arxiv.org/abs/1406.0341}{{\ttfamily
  1406.0341}}].

\bibitem{Hebecker:2018fln}
A.~Hebecker, D.~Junghans and A.~Schachner, \emph{{Large Field Ranges from
  Aligned and Misaligned Winding}},
  \href{https://doi.org/10.1007/JHEP03(2019)192}{\emph{JHEP} {\bfseries 03}
  (2019) 192} [\href{https://arxiv.org/abs/1812.05626}{{\ttfamily
  1812.05626}}].

\bibitem{Palti:2015xra}
E.~Palti, \emph{{On Natural Inflation and Moduli Stabilisation in String
  Theory}}, \href{https://doi.org/10.1007/JHEP10(2015)188}{\emph{JHEP}
  {\bfseries 10} (2015) 188}
  [\href{https://arxiv.org/abs/1508.00009}{{\ttfamily 1508.00009}}].

\bibitem{Angus:2021jpr}
S.~Angus, K.-S.~Choi and C.S.~Shin, \emph{{Aligned natural inflation in the
  Large Volume Scenario}},
  \href{https://doi.org/10.1007/JHEP10(2021)248}{\emph{JHEP} {\bfseries 10}
  (2021) 248} [\href{https://arxiv.org/abs/2106.09853}{{\ttfamily
  2106.09853}}].

\bibitem{Peloso:2015dsa}
M.~Peloso and C.~Unal, \emph{{Trajectories with suppressed tensor-to-scalar
  ratio in Aligned Natural Inflation}},
  \href{https://doi.org/10.1088/1475-7516/2015/06/040}{\emph{JCAP} {\bfseries
  06} (2015) 040} [\href{https://arxiv.org/abs/1504.02784}{{\ttfamily
  1504.02784}}].

\bibitem{Tristram:2021tvh}
M.~Tristram et~al., \emph{{Improved limits on the tensor-to-scalar ratio using
  BICEP and Planck data}},
  \href{https://doi.org/10.1103/PhysRevD.105.083524}{\emph{Phys. Rev. D}
  {\bfseries 105} (2022) 083524}
  [\href{https://arxiv.org/abs/2112.07961}{{\ttfamily 2112.07961}}].

\bibitem{Galloni:2022mok}
G.~Galloni, N.~Bartolo, S.~Matarrese, M.~Migliaccio, A.~Ricciardone and
  N.~Vittorio, \emph{{Updated constraints on amplitude and tilt of the tensor
  primordial spectrum}},
  \href{https://doi.org/10.1088/1475-7516/2023/04/062}{\emph{JCAP} {\bfseries
  04} (2023) 062} [\href{https://arxiv.org/abs/2208.00188}{{\ttfamily
  2208.00188}}].

\bibitem{CMB-S4:2020lpa}
{\scshape CMB-S4} collaboration, \emph{{CMB-S4: Forecasting Constraints on
  Primordial Gravitational Waves}},
  \href{https://doi.org/10.3847/1538-4357/ac1596}{\emph{Astrophys. J.}
  {\bfseries 926} (2022) 54}
  [\href{https://arxiv.org/abs/2008.12619}{{\ttfamily 2008.12619}}].

\bibitem{LiteBIRD:2023iei}
{\scshape LiteBIRD} collaboration, \emph{{Tensor-to-scalar ratio forecasts for
  extended LiteBIRD frequency configurations}},
  \href{https://doi.org/10.1051/0004-6361/202346155}{\emph{Astron. Astrophys.}
  {\bfseries 676} (2023) A42}
  [\href{https://arxiv.org/abs/2302.05228}{{\ttfamily 2302.05228}}].

\bibitem{Planck:2018vyg}
{\scshape Planck} collaboration, \emph{{Planck 2018 results. VI. Cosmological
  parameters}},
  \href{https://doi.org/10.1051/0004-6361/201833910}{\emph{Astron. Astrophys.}
  {\bfseries 641} (2020) A6}
  [\href{https://arxiv.org/abs/1807.06209}{{\ttfamily 1807.06209}}].

\bibitem{McAllister:2012am}
L.~McAllister, S.~Renaux-Petel and G.~Xu, \emph{{A Statistical Approach to
  Multifield Inflation: Many-field Perturbations Beyond Slow Roll}},
  \href{https://doi.org/10.1088/1475-7516/2012/10/046}{\emph{JCAP} {\bfseries
  10} (2012) 046} [\href{https://arxiv.org/abs/1207.0317}{{\ttfamily
  1207.0317}}].

\bibitem{Battefeld:2013xwa}
D.~Battefeld and T.~Battefeld, \emph{{A Smooth Landscape: Ending Saddle Point
  Inflation Requires Features to be Shallow}},
  \href{https://doi.org/10.1088/1475-7516/2013/07/038}{\emph{JCAP} {\bfseries
  07} (2013) 038} [\href{https://arxiv.org/abs/1304.0461}{{\ttfamily
  1304.0461}}].

\bibitem{Carta:2020oci}
F.~Carta, N.~Righi, Y.~Welling and A.~Westphal, \emph{{Harmonic hybrid
  inflation}}, \href{https://doi.org/10.1007/JHEP12(2020)161}{\emph{JHEP}
  {\bfseries 12} (2020) 161}
  [\href{https://arxiv.org/abs/2007.04322}{{\ttfamily 2007.04322}}].

\bibitem{Chatzistavrakidis:2012bb}
A.~Chatzistavrakidis, E.~Erfani, H.P.~Nilles and I.~Zavala, \emph{{Axiology}},
  \href{https://doi.org/10.1088/1475-7516/2012/09/006}{\emph{JCAP} {\bfseries
  09} (2012) 006} [\href{https://arxiv.org/abs/1207.1128}{{\ttfamily
  1207.1128}}].

\bibitem{Choi:2014rja}
K.~Choi, H.~Kim and S.~Yun, \emph{{Natural inflation with multiple
  sub-Planckian axions}},
  \href{https://doi.org/10.1103/PhysRevD.90.023545}{\emph{Phys. Rev. D}
  {\bfseries 90} (2014) 023545}
  [\href{https://arxiv.org/abs/1404.6209}{{\ttfamily 1404.6209}}].

\end{thebibliography}\endgroup

\end{document}